# Phyllosilicates as earth-abundant layered materials for electronics and optoelectronics: Prospects and challenges in their ultrathin limit

Ingrid D. Barcelos[1*], Raphaela de Oliveira[1,2], Gabriel R. Schleder[3], Matheus J. S. Matos[4], Raphael Longuinhos[5], Jenaina Ribeiro-Soares[5], Ana Paula M. Barboza[4], Mariana C. Prado[4], Elisângela S. Pinto[6], Yara Galvão Gobato[7], Hélio Chacham[2], Bernardo R. A. Neves[2], Alisson R. Cadore[3*]

[1]Brazilian Synchrotron Light Laboratory (LNLS), Brazilian Center for Research in Energy and Materials (CNPEM), 13083-100, Campinas, SP, Brazil

[2]Departamento de Física, Universidade Federal de Minas Gerais (UFMG), 31270-901, Belo Horizonte, MG, Brazil

[3]Brazilian Nanotechnology National Laboratory (LNNano), Brazilian Center for Research in Energy and Materials (CNPEM), 13083-100, Campinas, SP, Brazil

[4]Departamento de Física, Universidade Federal de Ouro Preto (UFOP), 35400-000, Ouro Preto, MG, Brazil

[5]Departamento de Física, Universidade Federal de Lavras (UFLA), 37200-900, Lavras, MG, Brazil.

[6]Campus Ouro Preto, Instituto Federal de Minas Gerais (IFMG), 35400-000, Ouro Preto, MG, Brazil

[7]Departamento de Física, Universidade Federal de São Carlos (UFSCar), 13565-905, São Carlos, SP, Brazil

*Corresponding authors: ingrid.barcelos@lnls.br; alisson.cadore@lnnano.cnpem.br

**ABSTRACT**

Phyllosilicate minerals are an emerging class of naturally occurring layered insulators with large bandgap energy that have gained attention from the scientific community. This class of lamellar materials has been recently explored at the ultrathin two-dimensional level due to their specific mechanical, electrical, magnetic, and optoelectronic properties, which are crucial for engineering novel devices (including heterostructures). Due to these properties, phyllosilicates minerals can be considered promising low-cost nanomaterials for future applications. In this Perspective article, we will present relevant features of these materials for their use in potential 2D-based electronic and optoelectronic applications, also discussing some of the major challenges in working with them.

## I. INTRODUCTION AND BACKGROUND

Ultrathin field effect transistors (FET), photodetectors, optical modulators and other multifunctional devices are building blocks for future electronic and optoelectronic applications.[1–3] Thus, aiming at novel nanotechnological products, the downscaling of these devices is required for miniaturized devices. In this scenario, Moore's law has shown exponential growth in FET number density with time. This inevitably requires that the length scales associated with the device get reduced. To satisfy Moore's law, substantial research efforts have been made to substitute or combine typical and well-established dielectrics with two-dimensional (2D) materials. This would hence reduce the sizes of the final devices.[1–3] The performance of these novel devices strongly depends not only on the properties of the active material, but also on the quality of the interface with the gate insulator (passive material). It is the insulating layer and its interface with the active material which decides the technological viability of a specific device.[2] Consequently, the right choice of insulating material preserves the optimum quality of the layered material (LM)-based device.[1–3]

Obtaining graphene using exfoliation techniques has opened a wide spectrum of research exploring van der Waals (vdWs) materials in novel 2D applications.[4,5] All these crystals are characterized by strong in-plane chemical bonds and weak out-of-plane vdWs forces, enabling their exfoliation down to a monolayer (1L).[5] The development of such 2D applications depends on the properties of LMs when they are reduced to few-layer (FL) samples, and when they are eventually combined into heterostructures.[6] In such a way, numerous new functional materials have been engineered by assembling LMs to build vertical vdWs heterostructures (vdWHs).[6–10]

Within the scope of electronic and optoelectronic integration, the miniaturization of vdWH-based devices is a promising route of appreciable commercial and scientific interest.[11–13] Atomically thin LMs can be obtained with bandgap energies that vary from zero to several electron volts (eV) and show strong light–matter interactions,[14,15] being well suited to act as building blocks for future technologies. The 2D-based devices' scalability demands sustainable and low-cost materials, especially in the case of insulating LMs, since the list of suitable alternative 2D insulators is currently very limited.[2] We highlight hexagonal boron nitride (hBN) as the main synthetic insulating LM explored in electronics and optoelectronic applications based on vdWHs.[2,16]

Although the engineering of nanodevices through synthetic routes can be beneficial in controlling specific properties, some routes can be non-trivial and may have a high cost associated. For this reason, large-scale production of these nanodevices may be challenging. In this context, phyllosilicate minerals emerge as attractive candidates for the next generation LMs.[17–19] Phyllosilicate minerals are wide bandgap insulators, stable in different environments, and abundant in nature.[17–27] They present a layered structure that enables their exfoliation down to 1L.[20,23,28,29] The variety of phyllosilicate specimens arises from ionic substitutions with different geometric arrangements of atoms within the crystal.[29,30] In this sense, the understanding of the general properties of phyllosilicates is required for the optimized applicability of these nanomaterials in future applications.

The atomic structures of phyllosilicates consist of the stacking of different layer types, as revealed in Fig. 1. Common to all phyllosilicates, the silicon oxide tetrahedral layer (T) forms a hexagonal lattice with basal oxygen (O) shared by silicon (Si) atoms in a ratio of 2 Si to 5 O, shown in Fig. 1, which also shows the genealogy of phyllosilicate mineral structures.[29,30] The general layered structure of phyllosilicate minerals is formed by stacking T layers intercalated by octahedral layers (Oc).[29] One unshared O per Si atom in the T layer points toward the Oc layer to form octahedrons with hydroxyl (OH) groups at the remaining vertices.[29] The Oc layers can be dioctahedral (gibbsite-like), with two octahedral sites to allocate trivalent ions, or trioctahedral (brucite-like), with three sites to allocate bivalent ions in a more compact arrangement.[29] A 1L-phyllosilicate can be formed by the combination of a T layer with an Oc layer in a T-Oc (1:1) or T-Oc-T (2:1) arrangement. Kaolinite and antigorite are standard specimens of 1:1 arrangement with dioctahedral and trioctahedral layers, respectively.[31] Phyllosilicates with 1:1 arrangement are more susceptible to polyhedral rotations and lattice distortions, which can form wave-like crystalline structures, such as serpentines.[31,32] Phyllosilicates with 2:1 arrangement present a compact structure with pyrophyllite and talc as standard specimens with dioctahedral and trioctahedral layers, respectively.[22,23,33] Structural differences between phyllosilicate specimens occur due to the different atomic substitutions in both T and Oc layers.[34] To maintain the charge balance of the structure, these atomic substitutions favor the formation of a cationic interlayer or even a hydroxide interlayer.[29] Micas are phyllosilicates pillared by cations in the interlayer space, while chlorites are phyllosilicates with the further formation of an octahedral hydroxide layer intercalating the T-Oc-T stacking.[29]

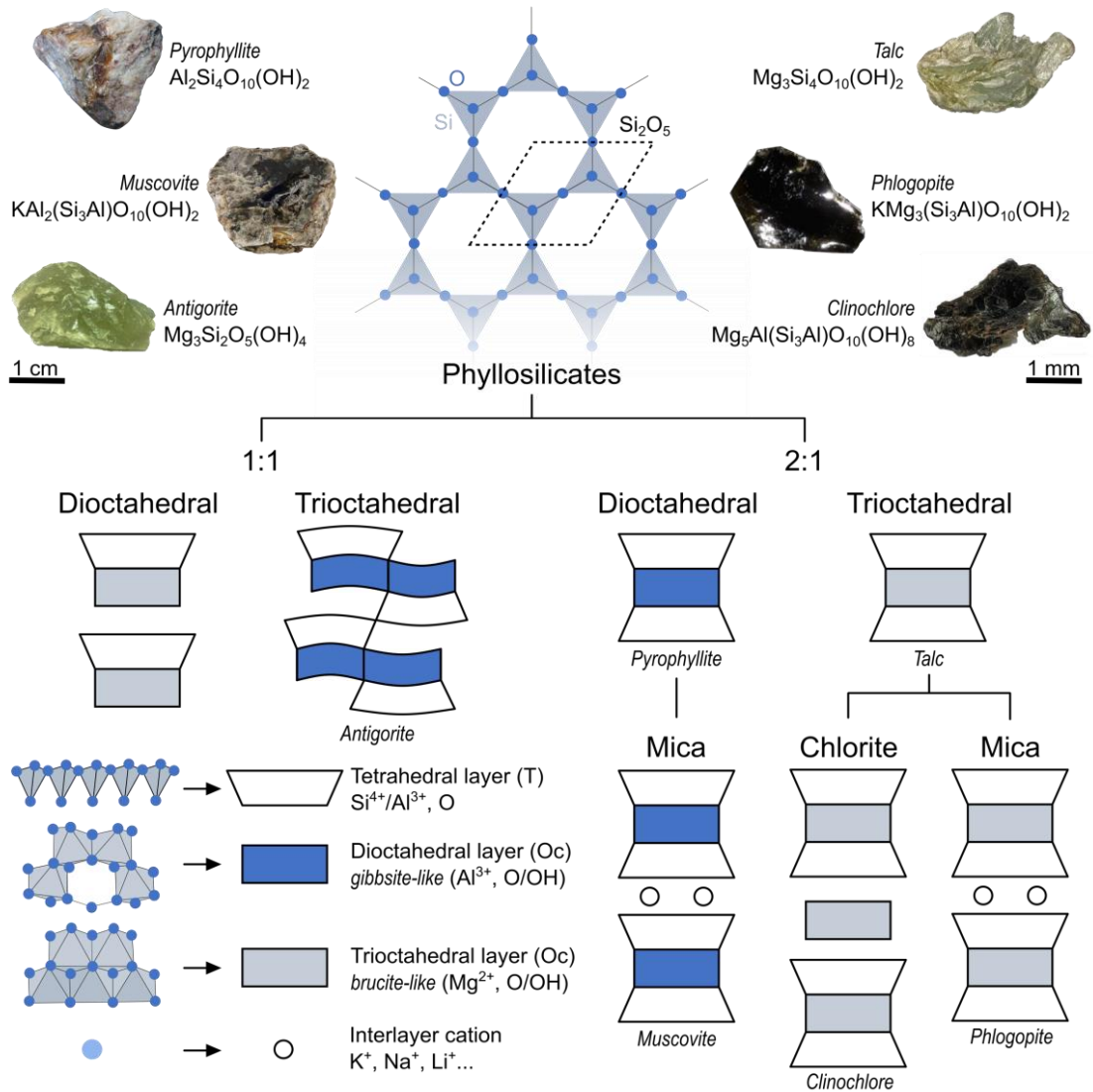

**FIG. 1.** Genealogical diagram of phyllosilicate mineral structures. The diagram starts by showing photographs of different specimens of phyllosilicates with a lamellar structure formed by a hexagonal lattice of Si-tetrahedrons with O at the vertices. Sequentially, phyllosilicate minerals can be divided into two classes according to the type of alternate stacking of tetrahedral (T) and octahedral (Oc) layers. The Oc layers can accommodate metallic ions in two or three sites. Typically, the dioctahedral layer is formed by trivalent ions, such as $Al^{3+}$, in the central position of octahedrons with O/OH at the vertices, while the trioctahedral layer is formed by bivalent ions as central atoms, such as $Mg^{2+}$. The variety of specimens occurs by ionic substitutions in the atomic structure of standard specimens. Mica subgroup is a direct alteration of the pyrophyllite-talc structure. In micas, interlayer cations emerge to balance the charge of the structure that is altered by ionic substitutions, such as one Al/Si substitution in the Oc layer. Chlorite subgroup is the most complex structure of phyllosilicate minerals in which the 2:1 layer is intercalated by a hydroxide layer, typically a brucite-like layer.

One of the challenges in using naturally occurring vdWs materials for 2D-based applications is the existence of impurities and defects in their crystal lattice.[2,20,22,27,29,35–37] There can be point defects (e.g., vacancies, substitutional or interstitial impurities) or extended defects (e.g., grain boundaries, twin planes and stacking faults), strongly affecting the electronic responses of the nanomaterial. Deep knowledge of all these features is essential to identify their eventual influence on a given property allowing their manipulation and tailoring to specific 2D technological applications. Moreover, the reduction in the dimensions of a system favors the emergence of quantum effects that can modify its bulk properties in surprising ways that might contribute to technological applications. As such, the advances of both computational strategies based on first-principles simulations and artificial intelligence (AI) techniques provide valuable theoretical complementary information towards the application of LMs.[38,39] These theoretical/computational techniques can be employed in all steps of the discovery and design workflow. Starting from the elucidation of the LMs structure to the prediction of their consequent properties, such as mechanical, vibrational, and electronic ones. Consequently, here, in this Perspective, we provide an overview of recent progress in theoretical and experimental studies of natural phyllosilicate minerals in multifunctional devices for electronics and optoelectronic applications. We first summarize the production of ultrathin phyllosilicates and the fabrication of vdWHs with other LMs in Sec. II. Next, their structural, magnetic, mechanical, and vibrational properties are discussed in Sec. III. We then illustrate their applications in electronics and optoelectronics in Sec. IV. Finally, section V provides the scientific and technological challenges and future directions of these earth abundant LMs.

## II. FABRICATION: FROM MECHANICAL EXFOLIATION AND STACKING TOWARDS LARGE AREA APPLICATIONS

### A. Mechanical Exfoliation and van der Waals Heterostructures

For all LMs to fulfill their potential to serve as a platform for new materials such as polymeric composites or multifunctional devices, it is essential to achieve large-scale production. Nevertheless, mechanical exfoliation using adhesive tape is a rapid, top-down proof-of-principle approach, and is an inexpensive method for testing whether a material yields FL-flakes easily. Several experimental techniques are used to verify that exfoliated LMs were not structurally damaged when compared to the parent bulk material.[21,22,31,35,40]

Mechanical exfoliation has been successfully applied to a variety of phyllosilicates, such as biotite,[37,40] serpentine,[31] phlogopite,[20] clinochlore,[27] talc[23–26,41–43], muscovite,[44–48] and others,[17,37] reaching stable single or FL-flakes onto different substrates. Thus, we foresee that novel vertical vdWHs will be made by manipulating (e.g., twisting, placing, and stacking) these phyllosilicates with other LMs to build unique artificial vdWHs with desired functionalities. The degree of freedom of aligned transfers suggests promising opportunities by freely tailoring the stacking angles between adjacent layers. This would remove, in principle, symmetry restrictions imposed by the thermodynamic stacking in natural crystals. However, we must comment that stacking such LMs by directly vdWs pick-up method can be challenging and polymer-assisted (e.g., PDMS or PMMA) transfer is a faster and easier approach to be applied, as shown in Refs.[42,49].

In general, layer-by-layer assembly provides a unique and versatile platform to (i) combine LMs in such a way as to translate their properties to macroscale devices and (2) innovate LM-based functional devices. These approaches allow the tuning of both thickness (number of transferred LMs) and hierarchical structure to achieve optimal performance metrics in a wide variety of applications. Moreover, achieving reliable transfer processes, suited for line integration of LMs and their vdWHs on top of integrated circuits, is strongly required because it would accelerate progress in optoelectronics and electronics as well as in photonics, and sensing applications. Additionally, besides integration protocols, industrial-scale applications may require other fabrication methods that can yield higher quantities of LMs.

### B. Liquid Phase Exfoliation

Liquid phase exfoliation (LPE) is a scalable and versatile route to produce thin nanoflakes of various LMs,[50,51] including non vdWs minerals.[52] For the sake of clarity, here LPE is considered a physical process that involves breaking bonds in materials through a mechanical energy source, resulting in nanostructures suspended in the exfoliation medium. It is distinct from chemical exfoliation, such as graphene oxide production, which involves chemical reactions and composition modifications of the processed material.[53] LPE has been successfully applied to a variety of LMs,[50–57] even if the starting material is not pure.[54] Thus, large-scale production of ultrathin phyllosilicates can be achieved by LPE.[51,57–59] Moreover, LPE can be combined with ion exchange, intercalation steps, or chemical modifications to aid the delamination of materials with stronger interlayer interactions, such as micas.[60–62] These additional steps can help achieving higher yield and thinner flakes. However, it is important to note that LPE typically results in a suspension of flakes with varying sizes and small lateral dimensions. Consequently, it is well-suited for generating nanomaterials used in applications such as mechanical reinforcement of composites and the production of conductive ink.[63] Given that, employing LPE to create insulating layers for electronic and optoelectronic devices poses a challenge due to the inherent characteristics of this technique.[64,65]

For non-expandable phyllosilicates, an intercalation or ion exchange step may be necessary to increase yield before performing LPE.[66] The behavior of minerals with surface charges differs depending on the charge site. While montmorillonite can be exfoliated in water, vermiculite does not exfoliate well even with long processing times.[66] Muscovite (a non-expandable clay) and biotite exposed to ultrasonic energy in water already display a drastic size reduction (from 2 mm to submicron sizes)[67] but even after 100 h of processing time are not reduced to FL-flakes. Intercalation of quaternary ammonium compounds is a commonly used strategy to increase interlayer space and delaminate the mineral. It also allows the modification of surface properties to make the material more compatible with hydrophobic polymers.[68]

Ion exchange is the most widely implemented method for preparing natural phyllosilicate mineral's ultrathin flakes.[66] Moreover, these cations between layers can also be used in energy harvesting.[69–71] Intercalation methods may be used to exfoliate minerals that do not have interchangeable ions or that the ions are strongly bound, such as kaolinite and illite.[66] Talc can be exfoliated using a variety of LPE setups. Simple acid medium sonication yields flakes with nanometer-scale thickness but not in the FL range.[72] To achieve FL-flakes, an appropriate medium choice is needed.[54,73] Saponite can be intercalated and exfoliated using an ultrasonic bath.[74] Polymers can be directly intercalated in organically modified silicates.[75,76] Montmorillonite is widely used to produce nanocomposites. It can be exfoliated by mechanical stirring and ultrasonication in different media, including water[66] and also can be intercalated in polymers by in situ polymerization.[77,78]

Finally, we must comment that synthetic phyllosilicates are also a promising route for production and have been produced for several decades.[79] Synthetic materials offer advantages such as high purity and controlled chemical composition[80] with respect to the natural obtained crystals. These silicates are usually sold by various companies and are typically synthesized using hydrothermal reactions. Macroscopic single crystals of fluorphlogopite can be synthetized using a batch melting technique with controlled cooling rate.[81] This method enables the growth of crystals typically exceeding several millimeters in size and with controlled composition. These techniques have been employed for several decades to produce materials utilized in petrology and cosmetics applications.[82] Furthermore, organoclays, which have widespread industrial applications, can be produced through the intercalation of organic molecules into the interlayer spaces of natural phyllosilicates.[83–87]

The fabrication methods to obtain nanoflakes described here are already employed to a variety of phyllosilicates. Nevertheless, they could be modified to exfoliate new minerals of that class or to tailor the product to the desired applications. Particularly, LPE and chemical exfoliation are under intense investigation to understand the basic science involved and to optimize the production towards industrial-scale applications. Furthermore, recent approaches based on machine learning (ML) can be employed to assist in different tasks related to obtaining large-scale samples. The simplest approach is to use ML to accelerate the identification of exfoliable materials.[88] Beyond this, the use of these techniques can assist in experimentally obtaining lateral-size control of exfoliated LMs.[89] In general, ML can assist in identifying experimental exfoliation conditions, similar to the case of chemical-vapor-deposition (CVD)-grown LMs.[90] A promising direction in this area is using natural language processing techniques of the literature data for discovering optimal experimental synthesis conditions.[91]

### III. PROPERTIES AND CHARACTERIZATION

#### A. Structural and Magnetic Properties

The different atomic structures of phyllosilicate crystals are shown in Fig. 1. Nevertheless, atomic-scale disorder due to substitutional impurities can be an intrinsic part of phyllosilicates and can strongly affect their fundamental properties (i.e., mechanical, vibrational, magnetic, and electronic). Within this class of LMs, we can identify Fe as the main source of point defects that are naturally present in those minerals.[20,27,29] As an abundant element in the geological formation environment of phyllosilicates,[92] Fe can be present as a substitutional impurity at different sites, coordination, and valence states simultaneously. A less expressive amount of other substitutional impurities, such as Mn, Ti and Cr, has already been reported in the literature.[20,27,35,93] However, it has been shown that Fe plays a crucial role in determining the macroscopic properties of phyllosilicates, such as magnetism[36,37], bandgap energy reduction[20,27,94] and vibrational fingerprints distortion.[92]

The magnetism that arises in phyllosilicates due to Fe impurities is highly dependent on the Fe concentration and the $Fe^{3+}/Fe^{2+}$ ratio.[36,37] Iron impurities can be present as substitutional ions in the $Fe^{3+}$ and $Fe^{2+}$ oxidation states at either four or six coordinated sites, the latter being the most common. The Oc layer is the main functional unit for magnetic ordering in naturally occurring Fe-rich phyllosilicates, such as minnesotaite, annite and biotite, which persists in FL.[37] Typically, phyllosilicates are paramagnetic at room temperature and can exhibit ferro or antiferromagnetic character below 40 K depending on the $Fe^{2+}$ content.[37,95] Reducing the total Fe content or increasing the $Fe^{3+}/Fe^{2+}$ ratio results in a stronger disorder that eventually destroys the magnetic ordering. It has already been reported in the literature that the critical temperature of phyllosilicates rises with increasing $Fe^{2+}$ fraction. Nevertheless, a small amount of $Fe^{3+}$ ions can also contribute to increase ferromagnetic interactions in phyllosilicates when it is surrounded by $Fe^{2+}$ ions due to exchange interactions. This trend establishes a possible route to tune the critical temperature of phyllosilicates and increase it significantly for magnetic applications.

Due to the complexity of phyllosilicate structure at the atomic level and the presence of several impurities, theoretical methods such as first-principles calculations are relevant to understanding their FL properties.[20,23,27,94] In the case of FL-talc, for instance, the first study on this LM involved a combination of atomic force microscopy (AFM) measurements and first-principles calculations within the density functional theory (DFT) to provide a description of its structural and mechanical properties, such as the Young and bending modulus.[23] For the clinochlore case, the disorder occurs intrinsically at the atomic level involving Al, Mg, Fe and Cr atoms, and these impurities affect the optical and structural properties of clinochlore,[27] as well as for other phyllosilicates.[20,36,37,94,96,97]

First-principles methods can be also an important tool in the explanation of experimental phenomena of phyllosilicate heterostructures involving distinct LMs. For instance, the spontaneous doping of graphene in graphene/talc vdWHs was theoretically identified as an indirect charge transfer effect arising from acceptor impurities in talc.[43] First-principles calculations of vibrational modes of FL-talc were also relevant to the explanation of the plasmon-phonon coupling in those vdWHs.[24] Consequently, an interesting challenge in this field would be a full first-principles description of the plasmon-phonon coupling, and the overall features of the effects imposed by different impurities types and amounts.

Given the increasing amount of theoretical and experimental data, ML methods can be employed to predict many structure-related properties, such as the essential requirement of thermodynamic stability.[98] With experimental data, it can be used for LMs identification with optical microscopy, Raman and photoluminescence (PL) spectroscopies, transmission electron microscopy, hyperspectral imaging, and AFM,[99–102] including the number of layers identification and real-time analysis. Towards realistic systems with defects, deep learning can assist in atomically resolving microscopy images for tracking defects local information,[103] as also designing optimal point defects for use in quantum computing for instance.[104] When creating doped systems via adsorption, data-driven design can assist with prediction of adsorption energies.[105] Towards increasing complexity with vdWHs, ML can accelerate their design from the many possible combinations to the ones with desired properties,[106] an active learning approach that minimizes the number of necessary samples[107,108] and that can even be employed in autonomous laboratories.[109]

### B. Mechanical Properties

Many theoretical and computational studies have been performed to investigate the mechanical properties of various LMs (e.g., bending, rolling, or stretching behavior). The mechanical properties of talc have already been determined using first-principles methods, and properties such as breaking strength ($\sigma_{max}$), elastic modulus, Poisson's ratio, and flexural stiffness were able to be obtained and compared with experimental results and other LMs.[23,26] From these calculations, it was possible to determine that talc has a $\sigma_{max}$ only 17% lower than graphene, the elastic modulus is approximately half of the value calculated for graphene, and 1L-talc is 30 times stiffer than graphene, with respect to flexural deformations. In another approach, the mechanical properties of clinochlore were also determined through hydrostatic compression and axial deformation analysis. Modeling of clinochlore allowed to obtain both the equation of state of this material and its second-order elastic constants, while the presence of Al impurities resulted in an increase in the strength of the material. These methods can be applied to determine other types of phyllosilicates, considering the interference of impurities in their structures on the mechanical properties.[110–112]

Structural and mechanical properties of FL-phyllosilicates at the atomic scale can be investigated from first principles. However, certain mechanical properties arise from scales much larger than interatomic distances. In this case, other theoretical techniques must be used. For instance, universal nanomechanical properties of suspended[113] and deposited[114] nanosheets of FL-phyllosilicates could only be theoretically analyzed through a combination of molecular dynamics and continuum-model methods. This indicates that future perspectives in this area might involve multi-scale theoretical investigations. The theoretical predictions of the mechanical properties of these LMs are valuable for electronic and optoelectronic applications because it may anticipate if a specific material from this family should be avoided for flexible devices on their ultrathin limit, for instance.

From an experimental perspective, measuring the nanomechanical properties of phyllosilicates is important for understanding their deformation mechanisms and interactions with other LMs. In this sense, scanning probe microscopy (SPM) is a powerful tool for characterizing the mechanical properties of nanomaterials. Its ability to provide quantitative data, with high spatial resolution makes it a useful technique for a wide range of applications in materials science, biology, and nanotechnology. The AFM cantilever deflection is a simple and, in general, nondestructive approach to obtain the elastic properties of atomically thin membranes. The force-distance curves obtained by AFM provide information on the elastic modulus, adhesive forces, and deformation behavior of the surfaces and have been used to measure graphene,[115] talc,[23] transition metal dichalcogenides (TMDs),[116] among others LMs.[117] Moreover, non-contact AFM can be employed to probe the distribution of $K^+$ atoms on cleaved mica.[118] Novel methodologies, based in 3D maps have been developed to perform nanomechanical characterization of materials. The basic principle of such methodology involves acquiring several force curves at every single pixel of the image. By monitoring the cantilever deflection, some mechanical properties of the sample can be determined with high spatial resolution across a sample surface, including their changes due to composition or structure variation. Another useful technique, from the SPM family, is the Scanning Tunneling Microscopy (STM), which is used to investigate surface topography and electronic properties of phyllosilicates with high resolution, providing further insights into the mechanical properties of these materials.[119]

Finally, ML techniques could be proved useful by analyzing structure-property relations for mechanical properties.[120] For instance, mechanical properties such as fracture strain, fracture strength, and Young's modulus can be correlated to the effects of chirality, system size, temperature, strain rate, and random vacancy defects.[121] Nanoscale friction can be understood using Bayesian modeling and transfer learning to predict the maximum energy barrier of the potential surface energy of LMs for optimizing tribological properties.[122] Beyond structural and defect visualization, deep learning can also enable strain mapping at single-atom resolution.[121]

### C. Vibrational Properties

The lattice vibrations in solids carry information on their chemical composition, chemical bonding - order and strength - and structure, which may depend on the thermodynamic conditions when the crystal was formed. In this Perspective, we summarize the usage of microscopic-based vibration spectroscopy techniques in phyllosilicates crystals and the very recent use of nanoscopic-based vibration spectroscopy techniques. The latter allows to surpass the diffraction limit of light, besides providing signal amplification, key factors for the study phyllosilicates nanoscopic samples.

The vibrational spectroscopy techniques (e.g., infrared (IR) and Raman spectroscopies), correlated with diffraction-based techniques (e.g., X-ray diffraction and electron-microprobe analyses), have unveiled the complexity of the atomic structure of phyllosilicates.[20,27,29,57,123,124] The spectroscopic signature of the phyllosilicates building blocks, i.e., the T and Oc units, the OH groups and, in some cases, the intercalant cations and layers between the T-Oc-T layers (see Fig. 1) were intensively studied.[92,125–131] The analysis reveals details on the interatomic forces and symmetries in these LMs; information about the wavenumber vibration window of these nanomaterials; as well as their vibration pattern and their relative peak intensity and width.[92,125–131]

The key factor for the success of the usage of vibrational spectroscopies in phyllosilicates is the difference in electronegativity between the atoms in their structure. This results in chemical bonds that display permanent dipole moments or with anisotropic polarizability under interaction with light. In this context, the strongest Raman scattering and IR absorption features will be related to vibrations of the tetrahedral units and OH/$H_2O$ groups, and minor features due to the vibration of the Oc modules. Yet, all the number of peaks, peak positions, relative intensities, and widths due to the vibration of these units and groups result in unique spectral features for each phyllosilicate.

Most of the literature on the vibrational spectra of silicate minerals concerns bulk samples and revealed a great richness of features, consistent with their overly complex structure. Some results could be rationalized, allowing the differentiation among silicate minerals.[92] Furthermore, the data allow the distinction between specimens of phyllosilicates minerals with gibbsite-like and brucite-like octahedral layers. For example, the analysis of a single feature can identify between dioctahedral or trioctahedral phyllosilicates.[92] In addition, within a given group of phyllosilicates, it may be possible to access the content of atomic substitutions (and impurities) in the T or Oc units that are characteristics of samples from different geological regions. It is worth noting the richness of effects and interactions due to the complexity of the structure of phyllosilicates revealed by the influence on their vibrations, as the ionic radius, masses of cations, related lattice distortion and crystal fields.[92,131–133] Yet, many challenges on the vibrational spectra-structural correlation remain due to the high complexity of the phyllosilicate structure and chemical variability in these LMs, which may vary for same minerals from different geological locations.

The Raman scattering in these phyllosilicates is quite challenging and the intensity of all Raman modes are very low. Consequently, it is almost impossible to investigate the complete Raman fingerprints of these LMs in their thinnest limit (down to 1L) by conventional systems. In this context, near-field optical microscopy (SNOM) techniques enabled the characterization of vibration modes in few to single layer crystals of conductors, semiconductors, and insulators LMs, as well as their vdWHs.[26,27,35,134–137] Besides surpassing the light diffraction limit, SNOM-based techniques for optical nanospectroscopy are capable of signal amplification. Moreover, in the case of synchrotron IR nanospectroscopy (SINS), out-of-plane modes are accessed in single-crystals due to the specific light-matter interaction through the tip-enhancement mechanism of signal.[24,26,35,137] Moreover, by combining point spectra with hyperspectral images with 25 nm resolution (typical tip size), the existence of impurities and defects could most likely be identified. Therefore, we foresee the use of more sophisticated SNOM-based tools in combination with first-principles calculations to assign the complete fingerprint of these LMs as a function of layers. Moreover, as we show next, this experimental technique demonstrates that vdWHs based on phyllosilicates and other LMs are relevant for novel optoelectronic applications in different spectral windows.

Note that by performing such IR and Raman measurements, besides their vibrational features, one can collect the low-frequency modes (<100 cm$^{-1}$) for estimation of the interlayer interactions and related mechanical properties of these LMs.[26,138,139] In addition to the aforementioned vibrational spectroscopies, Brillouin spectroscopy could be another useful technique to access the force constants between adjacent lamellas in phyllosilicates. This would also unveil their low-frequency rigid-layer modes. Although we could not find studies exploring this tool on LMs, and Ref. [140] presents some limitations of the current state-of-art technology in measuring LMs, we see it as an opportunity for the community to make future contributions. Furthermore, simulations, mostly based on DFT, are the main source of vibrational spectra for idealized reference materials, yielding systematic data potentially organized in databases. With the use of such data, vibrational properties such as vibrational stability[141] can be obtained from ML classifiers and used as a filtering and screening tool. Even these simulations can be accelerated using AI-learned atomistic potentials, producing the phonon spectra from which vibrational properties can be derived.[142]

## IV. ELECTRONIC AND OPTOELECTRONIC APPLICATIONS

Phyllosilicates show wide bandgap behavior, high dielectric constant, high thermal stability, are chemically inert and can be exfoliated down to 1L.[29,41,43,57,136,143] All these features result in a variety of possible electronic and optoelectronic applications. For example, Figs. 2(a)-(b) demonstrate the electrical strength of pyrophyllite[33] and muscovite[144] in their ultrathin limit, respectively. Similar experiments have been also performed in other phyllosilicates.[43,47,145] Fig. 2(c) presents the magnetic properties of Fe-rich talc down to 1L. Fig. 2(d) brings a mica-based resistive random-access memory (RRAM) based on the motion of $K^+$ ions under electric field, which could be applied in computing systems.[146] This effect has also been applied in muscovite/$WSe_2$ memory devices.[49] Moreover, Ref.[57] depicts different flexoelectric device designs for LPE-based-biotite material for energy generation and energy harvesting applications. Therefore, we visualize further studies using other phyllosilicate members for similar investigations.

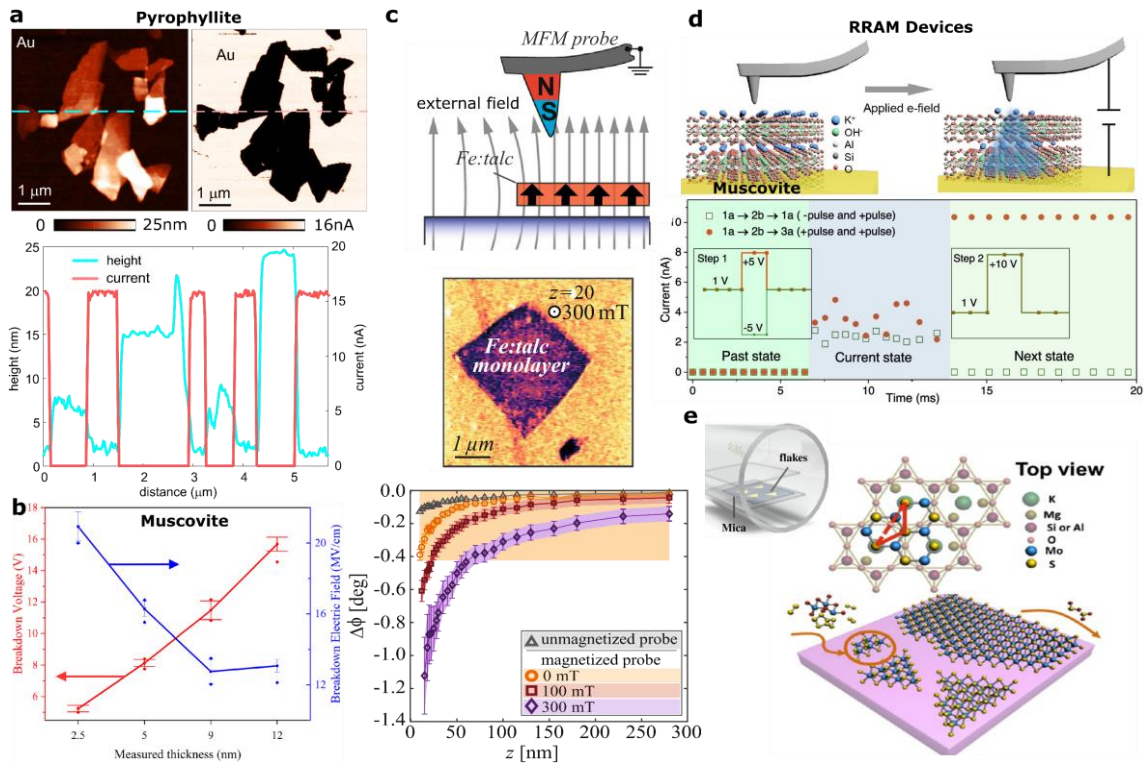

**FIG. 2**: Overview of 2D phyllosilicate crystals applications. a) Top panel: Topography and the corresponding current maps of the pyrophyllite flakes (obtained by LPE) deposited onto gold substrate. Bottom panel plots the overlapped height and current profiles along the dashed lines. Reproduced with permission from Appl Surf Sci **608**, 155114 (2023), Copyright 2022 Elsevier B.V. All rights reserved. b) Plot showing the thickness dependence of the voltage and electric field breakdown of ultrathin muscovite layers. A. Maruvada, K. Shubhakar, N. Raghavan, K.L. Pey, and S.J. O'Shea Sci Rep **12**(1), 14076 (2022) licensed under a Creative Commons Attribution (CC BY) license. c) Top panel: Sketch of the applied-field magnetic force microscopy (MFM) setup. Central panel shows the in-field (300 mT) second-pass phase lag map of a 1L-talc sample. Bottom panel brings the $\Delta\phi(z)$ for a 1L-talc considering non-magnetized probe (gray triangles), magnetized probe in 0 mT (orange circles), 100 mT (red squares), and 300 mT fields

(purple diamonds). A. Matković, L. Ludescher, O.E. Peil, A. Sharma, K.-P. Gradwohl, M. Kratzer, M. Zimmermann, J. Genser, D. Knez, E. Fisslthaler, C. Gammer, A. Lugstein, R.J. Bakker, L. Romaner, D.R.T. Zahn, F. Hofer, G. Salvan, J.G. Raith, and C. Teichert, NPJ 2D Mater Appl **5**(1), 94 (2021), licensed under a Creative Commons Attribution (CC BY) license. c) Top panel: schematic of the distribution of $K^+$ ions in the device before and after the application of a positive electric field. Bottom panel: Realization of a non-Markov chain in a single RRAM by a mica-based device. The three resistance states of the device are set by the applied pulse voltages. Zhang, W. Chen, C. Teng, W. Liao, B. Liu, and H.-M. Cheng, Sci Bull (Beijing) **66**(16), 1634–1640 (2021), licensed under a Creative Commons Attribution (CC BY) license. e) A schematic illustrating large-area growth of TMDs on mica. Reproduced with permission from Nano Lett **13**(8), 3870–3877 (2013). Copyright 2013 American Chemical Society.

It is important to recall that these phyllosilicates can be embedded into vdWHs, either by artificial stacking or growth routes. Therefore, the application portfolio of these LMs raises considerably. For example, phyllosilicate crystals (i.e., mica and fluorphlogopite) may be used as a growth substrate for other LMs[19,147–151] (see Fig. 2(e)). Moreover, ultrathin phyllosilicates crystals have been used as dielectric layers in a variety of compact 2D-based vdWHs devices, for instance in FETs,[42,152,153] memory devices,[49,146,153] and photodetectors[154–156] both on-chip and flexible substrates, Figs. 3(a)-(c). They have also been used as flexible substrates for organic light-emitting diodes,[157] but to the best of our knowledge not yet combined with other LMs for similar studies. Consequently, this is still open for further investigations.

Within the family of LMs, TMDs are also promising materials for the next generation of optoelectronics, spintronics devices, and quantum information technology. In addition, they can also be found in nature, such as $MoS_2$, $WS_2$, and others. The electronic quality of TMDs depends considerably on the incorporation of high-quality dielectric materials that possess a low defect density as well as being atomically smooth and uniform.[42,158–160] Previous studies investigated the electrical and optical properties of FET-based vdWHs with talc as a dielectric material for surface protection.[41,42] Although these investigations (combination of phyllosilicates and TMDs) are at their early stages compared to hBN-based TMDs vdWHs, there are great opportunities which include optimization of sample preparation and the use of several other phyllosilicate materials. Particularly, it was demonstrated that the full width at high maximum (FWHM) at low temperature of the exciton peak of 1L-$WS_2$/talc is comparable with 1L-$WS_2$/hBN.[42,161] This resulted in the observation of different excitonic complexes and showed that talc is indeed a promising dielectric to protect 1L-TMD from the $SiO_2$ substrate, Fig. 3(d).[42,161] Furthermore, the improvement of the optical quality of 1L-TMD on talc dielectric also contributed to uncovering further information on the valley physics of 1L-TMDs.[161] Additionally, it has been shown that the nano-PL emission of 1L-$MoS_2$/talc heterostructures shows sub-100nm wide charge oscillations due to different doping zones induced by talc layers, suggesting the ability to create ultrathin and ultrashort p-n homojunctions, Fig. 3(e).[136] Consequently, these previous results suggest that natural phyllosilicate materials could be a promising low-cost LM for vdWHs to explore fundamental physics and also for possible application in novel electronic and optoelectronic devices based on high-quality TMD heterostructures. Nevertheless, despite all advances in using such natural crystals in the 2D field, there are some challenges that need to be addressed in terms of phyllosilicate-based devices. These are mainly attributed to the impurity level acceptable for these LMs before losing their major characteristics or tailoring new features.

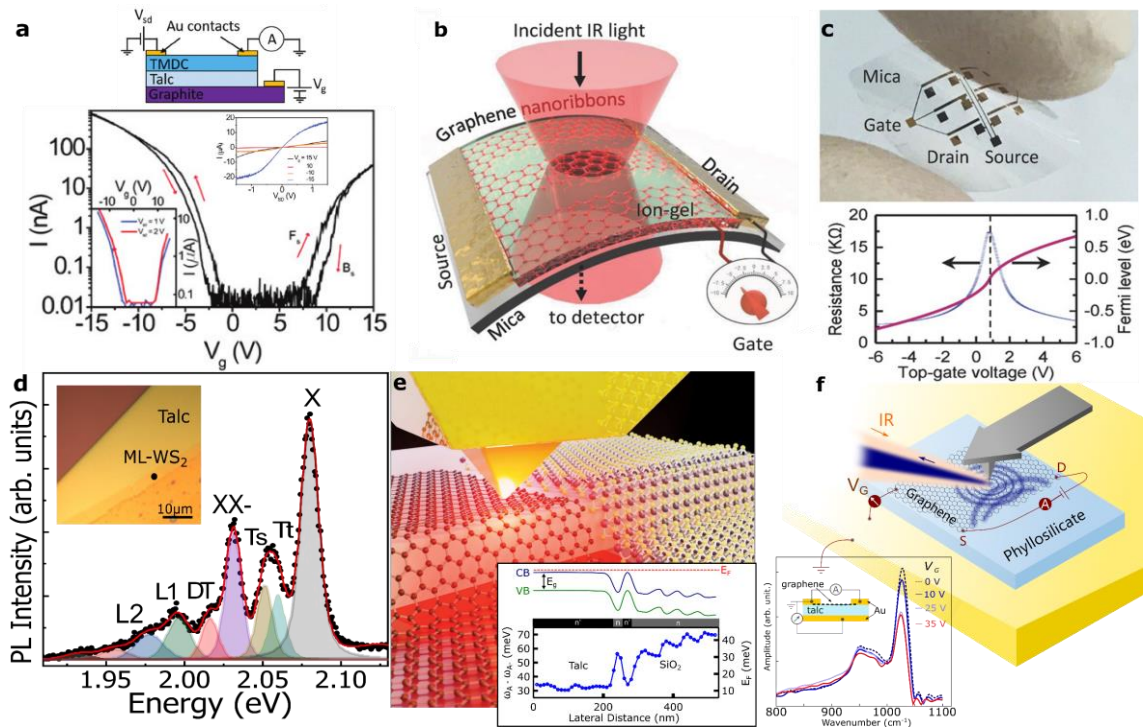

**FIG. 3.** Overview of 2D phyllosilicate crystals in vdWHs applications. (a) Schematic of the TMD/talc FET and representative transfer curve with $V_{SD}=0.03V$, where Fs and Bs represent the forwards and backwards sweeps, respectively. Insets: $IxVg$ for larger values of $V_{sd}$ and $I_{SD}xV_{SD}$ for select $Vg$ value. D. Nutting, G.A. Prando, M. Severijnen, I.D. Barcelos, S. Guo, P.C.M. Christianen, U. Zeitler, Y. Galvão Gobato, and F. Withers, Nanoscale **13**(37), 15853–15858 (2021), licensed under a Creative Commons Attribution (CC BY) license. (b) Flexible graphene/mica vdWHs plasmonic device. The figure shows a schematic illumination of plasmon excitation and detection of our flexible graphene/mica plasmonic device. (c) Photograph of the graphene/mica vdWHs plasmonic devices and typical transfer characteristics and Fermi level of graphene controlled by the ion–gel top gate. Figures (b-d) reproduced with permission from H. Hu, X. Guo, D. Hu, Z. Sun, X. Yang, and Q. Dai, Advanced Science **5**(8), 1800175 (2018), licensed under a Creative Commons Attribution (CC BY) license. (d) Optical microscope image and typical PL spectrum at 4.5 K of the 1L-WS$_2$/talc sample. Reproduced with permission from Phys Rev Appl **16**(6), 64055 (2021). Copyright 2021 by the American Physical Society. Near-field optical microscopy techniques (e) Tip enhanced spectroscopy and (f) s-SNOM to explore nano-optics in 2D-based materials. Figure (e) Reproduced with permission from J Phys Chem Lett **12**(31), 7625–7631 (2021). Copyright 2021 American Chemical Society. Figure (f) reproduced with permission from ACS Photonics **5**(5), 1912–1918 (2018). Copyright 2018, The American Chemical Society.

Although these natural systems are environmentally and mechanically stable and are free of dangling bonds, the impurities present in the crystal lattice may lead to a high and spontaneous charge transfer from the phyllosilicates to graphene (or TMDs) layers without any gate bias.[25,41–43,136] Hence, this effect could motivate the application of such ultrathin LMs as a steady doping tool both as substrate or passivation layer in cases where it is necessary to have a highly conductive (low sheet resistance) and stable graphene (or TMDs) layer as conducting material. For instance, in graphene solar cells with enhanced energy power conversion, a combination of steady doping and high-quality device is strongly desired. In addition, high doping on TMDs forming atomically thin p–n homojunction by employing TMD-compatible doping methods such as spontaneous charge transfer from the phyllosilicate substrates could be used to enhance the photovoltaic performance and energy harvesting in photovoltaic cells. Therefore, combining the spontaneous doping, high flexibility and mechanical strength of the phyllosilicate minerals would be by their own interesting for applying such low-cost materials in such energy applications,[57,162,163] in addition to the possibility of easy tailoring with 2D perovskites in novel optoelectronic applications.[164–166]

Natural phyllosilicate materials may have impurities such as Fe, Ni, or Co on their crystal structure, which can make them ferromagnetic (FM)/antiferromagnetic (aFM) LMs. Therefore, these materials could also be used in ultrathin vdWHs for possible application in spintronics. A recent study has shown that Fe-rich talc is air-stable and has evidenced a FM order at room temperature,[36] while another work demonstrated magnetic ordering in other Fe-rich phyllosilicates (i.e., minnesotaite, annite, and biotite).[37] These works could potentially motivate their use in vdWHs for spintronic applications. Actually, it has been already shown that 2D FM/aFM materials are promising building blocks for a new generation of spintronics devices.[167–169] Particularly, by combining FMs with 1L-TMDs, it should be possible to use magnetic proximity exchange effects for making materials magnetic without hosting magnetic ions.[170–173] This effect breaks the valley degeneracy which can be evidenced in circularly polarized resolved PL measurements without the use of high magnetic fields. In addition, the magnetic proximity exchange effects in 1L-TMDs due to stacked 2D FM materials can also be tuned by twisting the layers and by applying external gate voltages.[174] The interplay of exchange and valley Zeeman, and this gate and twist dependence considerably motivate the exploration of different vdWHs/devices with stacked TMDs and 2D magnets[167–169,174] which include the use of natural LMs. While there are several previous studies of vdWHs based on LM magnets, no such study combines natural magnetic phyllosilicate with 1L-TMDs. Therefore, different magnetic phyllosilicate materials could also be used to induce magnetic proximity effects and spin-dependent charge transfer in vdWHs based on 1L-TMDs.

We now bring attention to the use of natural phyllosilicates in the nanophotonics scope associated with experimental techniques based on nanoprobes, Fig. 3(f). It is known that plasmonic devices based on graphene (another natural LM) with electrical modulation are of great interest in diverse applications.[137,175–177] Their properties such as quality factors, electrical tunability, and lifetime strongly depend on the substrate.[178] Therefore, atomically flat insulating substrates with strong IR activity become very attractive for such novel devices. A recent study has shown that graphene plasmons' properties on flexible mica substrates present high-quality factors and a long lifetime.[179] More than a simple substrate, phyllosilicate nanocrystals have a strong IR activity from far- to mid-IR,[24,26,35,180] and for that reason, they are a natural platform for phonon-polaritons. Consequently, coupling between graphene plasmons and phyllosilicate phonon polaritons leads to hybrid plasmon–phonon polaritons. This has been studied in graphene/talc nanophotonic devices and it was shown that these hybrid modes can be finely controlled by external voltage bias (Fig. 3(f)).[24] This allows, therefore, the tailoring of complex nanophotonic devices based on such plasmonic systems. In this way, we foresee phyllosilicate materials as potential ultrathin photonic crystals whose nano-optical response can be tuned by device and thickness engineering. More importantly, their mechanical robustness, large bandgap, high flatness, and natural abundance potentially elect them and their vdWHs with diverse LMs (metal and semiconductors) to be a key element in low-cost ultracompact devices for polariton-based IR applications.

Finally, we must stress that there is a variety of available theoretical techniques for characterizing the electronic and optoelectronic properties of phyllosilicate minerals. First-principles electronic structure calculations, which provide information such as the band structure, projected density of states, optical absorption spectra, and phonons dispersions are undoubtedly significant approaches for complementing and enhancing the understanding of phyllosilicate properties. Theoretical methods can be used to model the presence of impurities in phyllosilicates in a robust way, thus assisting in the interpretation of experimental results. First-principles calculations can deal with all the structural complexity of this class of LMs. Thus, this tool is essential to understand and model physical phenomena not yet described for phyllosilicate minerals. Furthermore, with the use of data generated from these simulations and experiments is possible to predict many different electronic, optoelectronic, and more complex properties. ML is another tool that allows this task to be performed. The most direct application is the prediction of the electronic bandgap of the materials.[181,182] Targeting specific properties, and accelerated screening of photovoltaic materials can be made,[183] as such the in-silico discovery of electrocatalysts for the hydrogen evolution reaction and the oxygen evolution reaction[184] and efficient solar cell materials with adequate effective carrier masses.[185] Given small datasets but with sufficient data to encode relationships between properties and features, more complex phenomena can be assessed, such as semiconductor–metal transitions,[186] ferroelectric properties studied with unsupervised learning techniques,[187] magnetic ordering of LMs,[188] and topological insulators.[189] The evaluation of the low-temperature exciton valley polarization landscape can be made possible by using position-dependent PL information as input for Random Forests regression models, also revealing the feature input importance for understanding the spatially heterogeneous valley polarization.[190]

## V. CHALLENGES AND OUTLOOK

Although phyllosilicate minerals "were synthesized by mother nature" millions of years ago and over the last decades studied in the bulk form by several groups from different countries, only recently some of these LMs were investigated in the 2D form. Furthermore, while numerous works presented unique methods for obtaining FL-phyllosilicates and established their use in a plethora of novel applications, we believe that there are several aspects that warrant further investigations but have received less attention.

The atomically smooth and flexible interface of phyllosilicates provides additional freedom for combining different technological functions, allowing easy creation of more efficient and specific applications in hybrid systems.[1,19] Thus, these earth-abundant low-cost nanomaterials would allow downscaling of novel multifunctional devices as required for cheaper and miniaturized systems. However, novel architectures, designs and fabrication methods using these LMs in multifunctional 2D-based devices can still be challenging. Therefore, it is important to choose the right device structure and design for specific applications since these features strongly affect the overall device performance. Moreover, a common and central challenge in the study of these natural crystals is the variation of impurity concentration across the minerals in different earth locations with different types of impurities. Such variations in the impurity density can be a major source of disorder that needs to be carefully considered to fully understand the material´s properties (i.e., structural, electrical, and optical) and propose new technological applications.

For instance, depending on the amount of element substitutions in the crystal lattice, one material may show physical properties closer to one or another material. Biotite is an example of a solid-solution regarding Fe concentration between the annite – phlogopite and siderophyllite – eastonite endmembers that may show mixed properties of these specimens.[29] In addition, by changing the substitutional impurity type and amount (i.e., Fe, Ni, or Co) in the Mg or Al sites in the central Oc layer, phyllosilicate crystals may become magnetic vdWs materials and could serve as a versatile platform for 2D magnetic insulators.[36,191–193] Hence, they could be used to induce magnetic proximity effects in other LMs, such as TMDs or graphene layers. Likewise, phyllosilicate materials under certain impurity substitutions are also appealing because they can meet symmetry requirements for piezoelectricity,[194] as well as spontaneous charge transfer.[42,43,136] Additionally, the cations (usually $K^+$ and $Mg^{2+}$) adsorbed on the basal plane surfaces and residing in the space between the layers of these phyllosilicates can work as ion exchangers since they can be replaced by other ions (i.e., $H^+$, $Li^+$ or $Cs^+$), meaning that they may act as sponges absorbing and releasing such ions, and work in ion-selective transport and energy harvesting.[57,69–71] Consequently, there still exist a few outstanding challenges and barriers that should be taken into consideration in the further development of multifunctional devices based on natural and earth-abundant phyllosilicates 2D materials.

In this sense, ML is a powerful tool to overcome several of the presented challenges. Since one of the main challenges in the area of LMs, and materials science in general, is to solve the structure-properties relationship, with additional complexities of quantum confinement effects, most AI sub-fields can be employed to answer specific questions. These include ML itself, deep learning, natural language processing, computer vision, expert recommendation systems, and automation with robotics. The tools can aid the scientist in the screening process among all the phyllosilicate members to find the optimal candidates for specific applications or to investigate the impact of the impurities or other structural changes to a given property. In such cases, ML can predict directly physical properties, what is known as tertiary output in the context of DFT simulations,[195] or even encode the relation to learn the secondary outputs, which are total energies and atomic forces, clarifying the structure-property relationships. Such strategies, thus, allow going beyond the traditional design of materials with desired properties, accelerating this discovery using inverse design.[189] Going forward, the recent rapid development of the areas of computer vision, natural language processing, and ML-forcefields are obvious routes for further advances, benefiting respectively microscopy and spectral data analysis, literature insight extraction, and understanding of kinetics, dynamical processes, stability, extended defects, and interactions.[196] On a broader perspective, ML can even be employed to understand the cytotoxic behavior of materials, related to their lifecycle.[197]

We must also comment that in flexible devices and nano-electromechanical systems, the properties of phyllosilicates depend on a variety of factors, including the composition of the mineral, the orientation of the sheets, and the degree of interlayer bonding. In general, as a bulk form, phyllosilicates are relatively soft minerals with low hardness values. Typically, they can be easily scratched, bent, or twisted without breaking by harder materials. Consequently, phyllosilicate lubricants have gained attention as a potential alternative to conventional lubricants due to their unique layered structure and excellent lubrication properties.[21–23,31] Furthermore, recent studies have revealed that certain phyllosilicates such as talc, exhibit significant breaking strength and flexural rigidity at the nanoscale, along with moderate stiffness. Such LMs can withstand deformations with small curvature radii without tearing apart, despite being one order of magnitude stiffer than graphene.[23] Because of that, FL-talc are suitable for mechanical reinforcement in nanocomposites like ceramics or biomaterials, such as collagen or nanocellulose fibers, which is important to technological and biomedical developments in the future.[198] In addition, during mechanical manipulation (i.e., bending, twisting, and tapping) on these flexible nanomaterials, a flexural response inside the layers can give rise to polarization due to the rearrangement of surface charge in these ultrathin surface-charged LMs, hence a gain in the energy was measured in FL-biotite.[57] Consequently, this flexoelectric effect could be applied to many other ultrathin phyllosilicates and could potentially pave the way for next-generation energy harvesting through naturally obtained metal oxide minerals with energy harvesting properties.[163]

In conclusion, in this Perspective, we presented an overview of the current studies involving structural, electrical, magnetic, mechanical, and optoelectronic properties and application of earth-abundant phyllosilicate minerals, as depicted in Fig. 4. Thereafter, we believe that there are indeed great opportunities for future application of these naturally obtained LMs in vdWHs for electronics, optoelectronics, spintronics and nanophotonics.

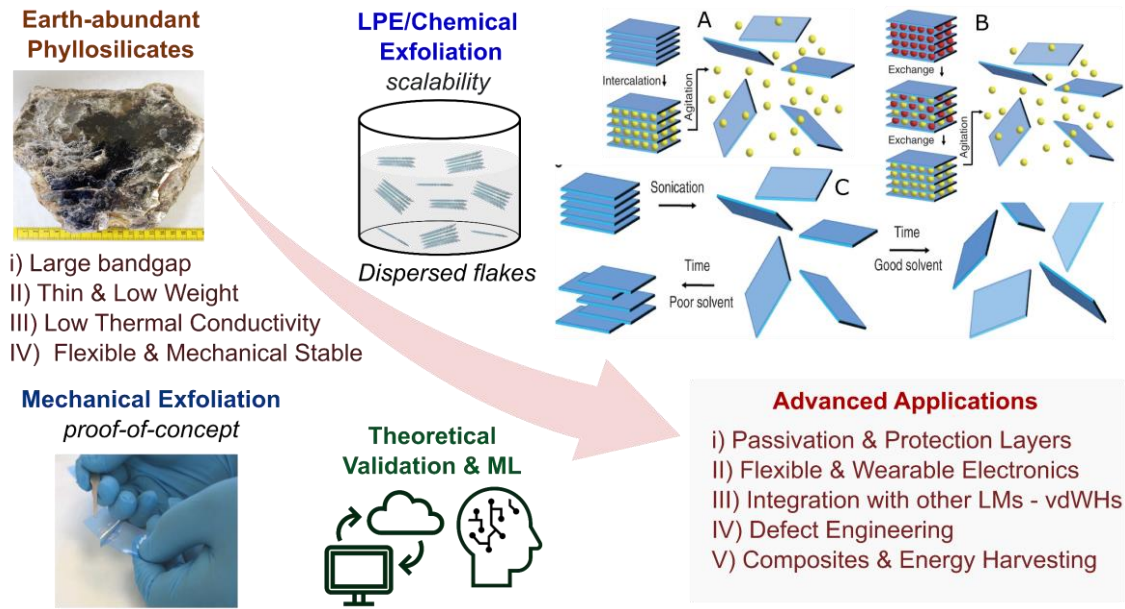

**FIG. 4.** Schematic of the roadmap of Earth-abundant phyllosilicate minerals from natural mines to advanced applications. The evolution process is heavily dependent on proof-of-concept studies using mechanical exfoliation processing, first-principles studies, and machine learning techniques. Thus, large-area applications can be envisioned by applying LPE and the different chemical exfoliation processes: (A) Ion intercalation; (B) Ion exchange; (C) Sonication assisted exfoliation. Figures (a-c) reproduced with permission from Science (1979) **340**(6139), 72–75 (2013). Copyright 2013, The American Association for the Advancement of Science.

**ACKNOWLEDGMENTS**

All authors thank the financial support from Coordination for the Improvement of Higher Education Personnel (CAPES), the National Council for Scientific and Technological Development (CNPq), Fundação de Amparo à Pesquisa do Estado de Minas Gerais (FAPEMIG), and the Brazilian Nanocarbon Institute of Science and Technology (INCT/Nanocarbono). M.J.S.M, A.P.M.B and M.C.P acknowledge the support from the Universidade Federal de Ouro Preto (UFOP). Y.G.G acknowledges the financial support of Fundação de Amparo a Pesquisa do Estado de São Paulo (FAPESP) (22/00419-0 and 22/08329-0). R.L and J.R.-S acknowledge the support from FAPEMIG (APQ-01553-22, APQ-01922-21, RED-00282–16, RED-00185–16), CNPq (408319/2021-6 and 312865/2020-1), and from the Universidade Federal de Lavras (UFLA).

**REFERENCES**

[1] M.C. Lemme, D. Akinwande, C. Huyghebaert, and C. Stampfer, "2D materials for future heterogeneous electronics," Nat Commun **13**(1), 2–6 (2022).

[2] Y.Y. Illarionov, T. Knobloch, M. Jech, M. Lanza, D. Akinwande, M.I. Vexler, T. Mueller, M.C. Lemme, G. Fiori, F. Schwierz, and T. Grasser, "Insulators for 2D nanoelectronics: the gap to bridge," Nat Commun **11**(1), 3385 (2020).

[3] X. Huang, C. Liu, and P. Zhou, "2D semiconductors for specific electronic applications: from device to system," NPJ 2D Mater Appl **6**(1), 51 (2022).

[4] C. Backes, A.M. Abdelkader, C. Alonso, A. Andrieux-Ledier, R. Arenal, J. Azpeitia, N. Balakrishnan, L. Banszerus, J. Barjon, R. Bartali, S. Bellani, C. Berger, R. Berger, M.M.B. Ortega, C. Bernard, P.H. Beton, A. Beyer, A. Bianco, P. Bøggild, F. Bonaccorso, G.B. Barin, C. Botas, R.A. Bueno, D. Carriazo, A. Castellanos-Gomez, M. Christian, A. Ciesielski, T. Ciuk, M.T. Cole, J. Coleman, C. Coletti, L. Crema, H. Cun, D. Dasler, D. De Fazio, N. Díez, S. Drieschner, G.S. Duesberg, R. Fasel, X. Feng, A. Fina, S. Forti, C. Galiotis, G. Garberoglio, J.M. García, J.A. Garrido, M. Gibertini, A. Gölzhäuser, J. Gómez, T. Greber, F. Hauke, A. Hemmi, I. Hernandez-Rodriguez, A. Hirsch, S.A. Hodge, Y. Huttel, P.U. Jepsen, I. Jimenez, U. Kaiser, T. Kaplas, H. Kim, A. Kis, K. Papagelis, K. Kostarelos, A. Krajewska, K. Lee, C. Li, H. Lipsanen, A. Liscio, M.R. Lohe, A. Loiseau, L. Lombardi, M. Francisca López, O. Martin, C. Martín, L. Martínez, J.A. Martin-Gago, J. Ignacio Martínez, N. Marzari, Á. Mayoral, J. McManus, M. Melucci, J. Méndez, C. Merino, P. Merino, A.P. Meyer, E. Miniussi, V. Miseikis, N. Mishra, V. Morandi, C. Munuera, R. Muñoz, H. Nolan, L. Ortolani, A.K. Ott, I. Palacio, V. Palermo, J. Parthenios, I. Pasternak, A. Patane, M. Prato, H. Prevost, V. Prudkovskiy, N. Pugno, T. Rojo, A. Rossi, P. Ruffieux, P. Samorì, L. Schué, E. Setijadi, T. Seyller, G.

Speranza, C. Stampfer, I. Stenger, W. Strupinski, Y. Svirko, S. Taioli, K.B.K. Teo, M. Testi, F. Tomarchio, M. Tortello, E. Treossi, A. Turchanin, E. Vazquez, E. Villaro, P.R. Whelan, Z. Xia, R. Yakimova, S. Yang, G.R. Yazdi, C. Yim, D. Yoon, X. Zhang, X. Zhuang, L. Colombo, A.C. Ferrari, and M. Garcia-Hernandez, "Production and processing of graphene and related materials," 2d Mater **7**(2), 022001 (2020).

[5] K.S. Novoselov, D. Jiang, F. Schedin, T.J. Booth, V. V Khotkevich, S. V Morozov, and A.K. Geim, "Two-dimensional atomic crystals," Proc Natl Acad Sci U S A **102**(30), 10451–10453 (2005).

[6] A.K. Geim, and I. V Grigorieva, "Van der Waals heterostructures.," Nature **499**(7459), 419–25 (2013).

[7] G. Wang, A. Chernikov, M.M. Glazov, T.F. Heinz, X. Marie, T. Amand, and B. Urbaszek, "Colloquium: Excitons in atomically thin transition metal dichalcogenides," Rev Mod Phys **90**(2), 21001 (2018).

[8] S. Hamza Safeer, A.S.M. V Ore, A.R. Cadore, V.O. Gordo, P.G. Vianna, I.C.S. Carvalho, V. Carozo, and C.J.S. de Matos, "CVD growth and optical characterization of homo and heterobilayer TMDs," J Appl Phys **132**(2), 024301 (2022).

[9] A.R. Cadore, E. Mania, K. Watanabe, T. Taniguchi, R.G. Lacerda, and L.C. Campos, "Thermally activated hysteresis in high quality graphene/h-BN devices," Appl Phys Lett **108**(23), 233101 (2016).

[10] M.S.G. Feuer, A.R.-P. Montblanch, M.Y. Sayyad, C.M. Purser, Y. Qin, E.M. Alexeev, A.R. Cadore, B.L.T. Rosa, J. Kerfoot, E. Mostaani, R. Kałęba, P. Kolari, J. Kopaczek, K. Watanabe, T. Taniguchi, A.C. Ferrari, D.M. Kara, S. Tongay, and M. Atatüre, "Identification of Exciton Complexes in Charge-Tunable Janus WSeS Monolayers," ACS Nano **17**(8), 7326–7334 (2023).

[11] J. Cheng, C. Wang, X. Zou, and L. Liao, "Recent Advances in Optoelectronic Devices Based on 2D Materials and Their Heterostructures," Adv Opt Mater **7**(1), 1800441 (2019).

[12] J. An, X. Zhao, Y. Zhang, M. Liu, J. Yuan, X. Sun, Z. Zhang, B. Wang, S. Li, and D. Li, "Perspectives of 2D Materials for Optoelectronic Integration," Adv Funct Mater **32**(14), 2110119 (2022).

[13] J. He, I. Paradisanos, T. Liu, A.R. Cadore, J. Liu, M. Churaev, R.N. Wang, A.S. Raja, C. Javerzac-Galy, P. Roelli, D. De Fazio, B.L.T. Rosa, S. Tongay, G. Soavi, A.C. Ferrari, and T.J. Kippenberg, "Low-Loss Integrated Nanophotonic Circuits with Layered Semiconductor Materials," Nano Lett **21**(7), 2709–2718 (2021).

[14] T. Mueller, and E. Malic, "Exciton physics and device application of two-dimensional transition metal dichalcogenide semiconductors," NPJ 2D Mater Appl **2**(1), 1–12 (2018).

[15] A. Chaves, J.G. Azadani, H. Alsalman, D.R. da Costa, R. Frisenda, A.J. Chaves, S.H. Song, Y.D. Kim, D. He, J. Zhou, A. Castellanos-Gomez, F.M. Peeters, Z. Liu, C.L. Hinkle, S.-H. Oh, P.D. Ye, S.J. Koester, Y.H. Lee, Ph. Avouris, X. Wang, and T. Low, "Bandgap engineering of two-dimensional semiconductor materials," NPJ 2D Mater Appl **4**(1), 29 (2020).

[16] T. Knobloch, Y.Y. Illarionov, F. Ducry, C. Schleich, S. Wachter, K. Watanabe, T. Taniguchi, T. Mueller, M. Waltl, M. Lanza, M.I. Vexler, M. Luisier, and T. Grasser, "The performance limits of hexagonal boron nitride as an insulator for scaled CMOS devices based on two-dimensional materials," Nat Electron **4**(2), 98–108 (2021).

[17] R. Frisenda, Y. Niu, P. Gant, M. Muñoz, and A. Castellanos-Gomez, "Naturally occurring van der Waals materials," NPJ 2D Mater Appl **4**(1), 1–13 (2020).

[18] Z. Li, and Derek. Wu, *Mica: Properties, Synthesis and Applications* (Nova Science Publishers, Hauppauge, N.Y, 2012).

[19] Y. Bitla, and Y.-H. Chu, "MICAtronics: A new platform for flexible X-tronics," FlatChem **3**, 26–42 (2017).

[20] A.R. Cadore, R. de Oliveira, R. Longuinhos, V. de C. Teixeira, D.A. Nagaoka, V.T. Alvarenga, J. Ribeiro-Soares, K. Watanabe, T. Taniguchi, R.M. Paniago, A. Malachias, K. Krambrock, I.D. Barcelos, and C.J.S. de Matos, "Exploring the structural and optoelectronic properties of natural insulating phlogopite in van der Waals heterostructures," 2d Mater **9**(3), 035007 (2022).

[21] B. Vasic, C. Czibula, M. Kratzer, B.R.A. Neves, A. Matkovic, and C. Teichert, "Two-dimensional talc as a van der Waals material for solid lubrication at the nanoscale," Nanotechnology **127**(1), 69–73 (2021).

[22] B. Vasić, R. Gajić, I. Milošević, Ž. Medić, M. Blagojev, M. Opačić, A. Kremenović, and D. Lazić, "Natural two-dimensional pyrophyllite: Nanoscale lubricant, electrical insulator and easily-machinable material," Appl Surf Sci, 155114 (2022).

[23] A.B. Alencar, A.P.M. Barboza, B.S. Archanjo, H. Chacham, and B.R.A. Neves, "Experimental and theoretical investigations of monolayer and fewlayer talc," 2d Mater **2**(1), 015004 (2015).

[24] I.D. Barcelos, A.R. Cadore, A.B. Alencar, F.C.B. Maia, E. Mania, R.F. Oliveira, C.C.B. Bufon, Â. Malachias, R.O. Freitas, R.L. Moreira, and H. Chacham, "Infrared Fingerprints of Natural 2D Talc and Plasmon–Phonon Coupling in Graphene–Talc Heterostructures," ACS Photonics **5**(5), 1912–1918 (2018).

[25] A.R. Cadore, E. Mania, A.B. Alencar, N.P. Rezende, S. de Oliveira, K. Watanabe, T. Taniguchi, H. Chacham, L.C. Campos, and R.G. Lacerda, "Enhancing the response of NH3 graphene-sensors by using devices with different graphene-substrate distances," Sens Actuators B Chem **266**, 438–446 (2018).

[26] R. Longuinhos, A.R. Cadore, H.A. Bechtel, C. J S De Matos, R.O. Freitas, J. Ribeiro-Soares, and I.D. Barcelos, "Raman and Far-Infrared Synchrotron Nanospectroscopy of Layered Crystalline Talc: Vibrational Properties, Interlayer Coupling, and Symmetry Crossover," The Journal of Physical Chemistry C **127**(12), 5876–5885 (2023).

[27] R. de Oliveira, L.A.G. Guallichico, E. Policarpo, A.R. Cadore, R.O. Freitas, F.M.C. da Silva, V. de C. Teixeira, R.M. Paniago, H. Chacham, M.J.S. Matos, A. Malachias, K. Krambrock, and I.D. Barcelos, "High

throughput investigation of an emergent and naturally abundant 2D material: Clinochlore," Appl Surf Sci **599**, 153959 (2022).

28 F. Jia, and S. Song, "Preparation of monolayer muscovite through exfoliation of natural muscovite," RSC Adv **5**(65), 52882–52887 (2015).

29 W.A. Deer, R.A. Howie, and J. Zussman, *An Introduction to the Rock-Forming Minerals* (Mineralogical Society of Great Britain and Ireland, 2013).

30 Jiri Konta, "Phyllosilicates in the sediment-forming processes: weathering, erosion, transportation, and deposition," Acta Geodynamica et Geomaterialia **6**(153), 13–43 (2009).

31 J.C.C. Santos, A.P.M. Barboza, M.J.S. Matos, I.D. Barcelos, T.F.D. Fernandes, E.A. Soares, R.L. Moreira, and B.R.A. Neves, "Exfoliation and characterization of a two-dimensional serpentine-based material," Nanotechnology **30**(44), 445705 (2019).

32 X. Li, Q. Liu, H. Cheng, S. Zhang, and R.L. Frost, "Mechanism of kaolinite sheets curling via the intercalation and delamination process," J Colloid Interface Sci **444**, 74–80 (2015).

33 B. Vasić, R. Gajić, I. Milošević, Ž. Medić, M. Blagojev, M. Opačić, A. Kremenović, and D. Lazić, "Natural two-dimensional pyrophyllite: Nanoscale lubricant, electrical insulator and easily-machinable material," Appl Surf Sci **608**, 155114 (2023).

34 W.F. Bleam, "The Nature of Cation-Substitution Sites in Phyllosilicates," Clays Clay Miner **38**(5), 527–536 (1990).

35 R. de Oliveira, A.R. Cadore, R.O. Freitas, and I.D. Barcelos, "Review on infrared nanospectroscopy of natural 2D phyllosilicates," Journal of the Optical Society of America A **40**(4), C157–C168 (2023).

36 A. Matković, L. Ludescher, O.E. Peil, A. Sharma, K.-P. Gradwohl, M. Kratzer, M. Zimmermann, J. Genser, D. Knez, E. Fisslthaler, C. Gammer, A. Lugstein, R.J. Bakker, L. Romaner, D.R.T. Zahn, F. Hofer, G. Salvan, J.G. Raith, and C. Teichert, "Iron-rich talc as air-stable platform for magnetic two-dimensional materials," NPJ 2D Mater Appl **5**(1), 94 (2021).

37 M.Z. Khan, O.E. Peil, A. Sharma, O. Selyshchev, S. Valencia, F. Kronast, M. Zimmermann, M.A. Aslam, J.G. Raith, C. Teichert, D.R.T. Zahn, G. Salvan, and A. Matković, "Probing Magnetic Ordering in Air Stable Iron-Rich Van der Waals Minerals," Advanced Physics Research **2300070**, 1–13 (2023).

38 G.R. Schleder, A.C.M. Padilha, C.M. Acosta, M. Costa, and A. Fazzio, "From DFT to machine learning: recent approaches to materials science–a review," Journal of Physics: Materials **2**(3), 032001 (2019).

39 G.R. Schleder, A.C.M. Padilha, A. Reily Rocha, G.M. Dalpian, and A. Fazzio, "Ab Initio Simulations and Materials Chemistry in the Age of Big Data," J Chem Inf Model **60**(2), 452–459 (2020).

40 W. Bower, W. Head, G.T.R. Droop, R. Zan, R.A.D. Pattrick, P. Wincott, and S.J. Haigh, "High-resolution imaging of biotite using focal series exit wavefunction restoration and the graphene mechanical exfoliation method," Mineral Mag **79**(2), 337–344 (2015).

41 G.A. Prando, M.E. Severijnen, I.D. Barcelos, U. Zeitler, P.C.M. Christianen, F. Withers, and Y. Galvão Gobato, "Revealing Excitonic Complexes in Monolayer WS2 on Talc Dielectric," Phys Rev Appl **16**(6), 64055 (2021).

42 D. Nutting, G.A. Prando, M. Severijnen, I.D. Barcelos, S. Guo, P.C.M. Christianen, U. Zeitler, Y. Galvão Gobato, and F. Withers, "Electrical and optical properties of transition metal dichalcogenides on talc dielectrics," Nanoscale **13**(37), 15853–15858 (2021).

43 E. Mania, A.B. Alencar, A.R. Cadore, B.R. Carvalho, K. Watanabe, T. Taniguchi, B.R.A. Neves, H. Chacham, and L.C. Campos, "Spontaneous doping on high quality talc-graphene-hBN van der Waals heterostructures," 2d Mater **4**(3), 031008 (2017).

44 A.P.M. Barboza, H. Chacham, C.K. Oliveira, T.F.D. Fernandes, E.H.M. Ferreira, B.S. Archanjo, R.J.C. Batista, A.B. de Oliveira, and B.R.A. Neves, "Dynamic Negative Compressibility of Few-Layer Graphene, h-BN, and MoS2," Nano Lett **12**(5), 2313–2317 (2012).

45 J. Shim, C.H. Lui, T.Y. Ko, Y.J. Yu, P. Kim, T.F. Heinz, and S. Ryu, "Water-gated charge doping of graphene induced by mica substrates," Nano Lett **12**(2), 648–654 (2012).

46 A. Castellanos-Gomez, M. Wojtaszek, N. Tombros, N. Agraït, B.J. Van Wees, and G. Rubio-Bollinger, "Atomically thin mica flakes and their application as ultrathin insulating substrates for graphene," Small **7**(17), 2491–2497 (2011).

47 M.R. Islam, and M. Tomitori, "Evaluation of the discrete thickness of exfoliated artificially synthesized mica nanosheets on silicon substrates: Toward characterization of the tunneling current through the nanosheets," Appl Surf Sci **532**(March), 147388 (2020).

48 J. Mohrmann, K. Watanabe, T. Taniguchi, and R. Danneau, "Persistent hysteresis in graphene-mica van der Waals heterostructures," Nanotechnology **26**(1), 015202 (2015).

49 Z. Yang, D. Wang, S. Wang, C. Tan, L. Yang, and Z. Wang, "Electrically Tuning Interfacial Ion Redistribution for mica/WSe2 Memory Transistor," Adv Electron Mater **9**(1), (2023).

50 F. Bonaccorso, A. Lombardo, T. Hasan, Z. Sun, L. Colombo, and A.C. Ferrari, "Production and processing of graphene and 2d crystals," Materials Today **15**(12), 564–589 (2012).

51 V. Nicolosi, M. Chhowalla, M.G. Kanatzidis, M.S. Strano, and J.N. Coleman, "Liquid exfoliation of layered materials," Science (1979) **340**(6139), 72–75 (2013).

52 A. Puthirath Balan, S. Radhakrishnan, C.F. Woellner, S.K. Sinha, L. Deng, C. de los Reyes, B.M. Rao, M. Paulose, R. Neupane, A. Apte, V. Kochat, R. Vajtai, A.R. Harutyunyan, C.-W. Chu, G. Costin, D.S. Galvao,

A.A. Martí, P.A. van Aken, O.K. Varghese, C.S. Tiwary, A. Malie Madom Ramaswamy Iyer, and P.M. Ajayan, "Exfoliation of a non-van der Waals material from iron ore hematite," Nat Nanotechnol **13**(7), 602–609 (2018).

[53] C. Backes, T.M. Higgins, A. Kelly, C. Boland, A. Harvey, D. Hanlon, and J.N. Coleman, "Guidelines for Exfoliation, Characterization and Processing of Layered Materials Produced by Liquid Exfoliation," Chemistry of Materials **29**(1), 243–255 (2017).

[54] A. Harvey, J.B. Boland, I. Godwin, A.G. Kelly, B.M. Szydłowska, G. Murtaza, A. Thomas, D.J. Lewis, P. O'Brien, and J.N. Coleman, "Exploring the versatility of liquid phase exfoliation: producing 2D nanosheets from talcum powder, cat litter and beach sand," 2d Mater **4**(2), 025054 (2017).

[55] I. Janica, S. Del Buffa, A. Mikołajczak, M. Eredia, D. Pakulski, A. Ciesielski, and P. Samorì, "Thermal insulation with 2D materials: Liquid phase exfoliated vermiculite functional nanosheets," Nanoscale **10**(48), 23182–23190 (2018).

[56] Y. Niu, J. Villalva, R. Frisenda, G. Sanchez-Santolino, L. Ruiz-González, E.M. Pérez, M. García-Hernández, E. Burzurí, and A. Castellanos-Gomez, "Mechanical and liquid phase exfoliation of cylindrite: a natural van der Waals superlattice with intrinsic magnetic interactions," 2d Mater **6**(3), 035023 (2019).

[57] P.L. Mahapatra, R. Tromer, P. Pandey, G. Costin, B. Lahiri, K. Chattopadhyay, A. P. M., A.K. Roy, D.S. Galvao, P. Kumbhakar, and C.S. Tiwary, "Synthesis and Characterization of Biotene: A New 2D Natural Oxide From Biotite," Small **18**(27), 2201667 (2022).

[58] J.N. Coleman, M. Lotya, A. O'Neill, S.D. Bergin, P.J. King, U. Khan, K. Young, A. Gaucher, S. De, R.J. Smith, I. V. Shvets, S.K. Arora, G. Stanton, H.-Y. Kim, K. Lee, G.T. Kim, G.S. Duesberg, T. Hallam, J.J. Boland, J.J. Wang, J.F. Donegan, J.C. Grunlan, G. Moriarty, A. Shmeliov, R.J. Nicholls, J.M. Perkins, E.M. Grieveson, K. Theuwissen, D.W. McComb, P.D. Nellist, and V. Nicolosi, "Two-Dimensional Nanosheets Produced by Liquid Exfoliation of Layered Materials," Science (1979) **331**(6017), 568–571 (2011).

[59] H. Chacham, J.C.C. Santos, F.G. Pacheco, D.L. Silva, R.M. Martins, J.P. Del'Boccio, E.M. Soares, R. Altoé, C.A. Furtado, F. Plentz, B.R.A. Neves, and L.G. Cançado, "Controlling the Morphology of Nanoflakes Obtained by Liquid-Phase Exfoliation: Implications for the Mass Production of 2D Materials," ACS Appl Nano Mater **3**(12), 12095–12105 (2020).

[60] X.-F. Pan, H.-L. Gao, Y. Lu, C.-Y. Wu, Y.-D. Wu, X.-Y. Wang, Z.-Q. Pan, L. Dong, Y.-H. Song, H.-P. Cong, and S.-H. Yu, "Transforming ground mica into high-performance biomimetic polymeric mica film," Nat Commun **9**(1), 2974 (2018).

[61] H.J. Bae, Y. Goh, H. Yim, S.Y. Yoo, J.-W. Choi, and D.-K. Kwon, "Atomically thin, large area aluminosilicate nanosheets fabricated from layered clay minerals," Mater Chem Phys **221**, 168–177 (2019).

[62] L. Weng-Lip, N.M. Salleh, N. 'Aqilah A. Rahman, N.S.A.A. Bakhtiar, H.M. Akil, and S.A. Zubir, "Enhanced intercalation of organo-muscovite prepared via hydrothermal reaction at low temperature," Bulletin of Materials Science **42**(5), 242 (2019).

[63] L. Niu, J.N. Coleman, H. Zhang, H. Shin, M. Chhowalla, and Z. Zheng, "Production of Two-Dimensional Nanomaterials via Liquid-Based Direct Exfoliation," Small **12**(3), 272–293 (2016).

[64] S. Akhavan, A. Ruocco, G. Soavi, A. Taheri Najafabadi, S. Mignuzzi, S. Doukas, A.R. Cadore, Y.A.K. Samad, L. Lombardi, K. Dimos, I. Paradisanos, J.E. Muench, H.F.Y. Watson, S. Hodge, L.G. Occhipinti, E. Lidorikis, I. Goykhman, and A.C. Ferrari, "Graphene-black phosphorus printed photodetectors," 2d Mater **10**(3), 035015 (2023).

[65] G. Hu, J. Kang, L.W.T. Ng, X. Zhu, R.C.T. Howe, C.G. Jones, M.C. Hersam, and T. Hasan, "Functional inks and printing of two-dimensional materials," Chem Soc Rev **47**(9), 3265–3300 (2018).

[66] K. Saha, J. Deka, R.K. Gogoi, K.K.R. Datta, and K. Raidongia, "Applications of Lamellar Membranes Reconstructed from Clay Mineral-Based Nanosheets: A Review," ACS Appl Nano Mater **5**(11), 15972–15999 (2022).

[67] L.A. Pérez-Maqueda, F. Franco, M.A. Avilés, J. Poyato, and J.L. Pérez-Rodríguez, "EFFECT OF SONICATION ON PARTICLE-SIZE DISTRIBUTION IN NATURAL MUSCOVITE AND BIOTITE," Clays Clay Miner **51**(6), 701–708 (2003).

[68] N.H. Che Ismail, N.S.A. Ahmad Bakhtiar, and H. Md. Akil, "Effects of cetyltrimethylammonium bromide (CTAB) on the structural characteristic of non-expandable muscovite," Mater Chem Phys **196**, 324–332 (2017).

[69] X. Wang, W. Tong, Y. Li, Z. Wang, Y. Chen, X. Zhang, X. Wang, and Y. Zhang, "Mica-based triboelectric nanogenerators for energy harvesting," Appl Clay Sci **215**, 106330 (2021).

[70] Y.-C. Zou, L. Mogg, N. Clark, C. Bacaksiz, S. Milovanovic, V. Sreepal, G.-P. Hao, Y.-C. Wang, D.G. Hopkinson, R. Gorbachev, S. Shaw, K.S. Novoselov, R. Raveendran-Nair, F.M. Peeters, M. Lozada-Hidalgo, and S.J. Haigh, "Ion exchange in atomically thin clays and micas," Nat Mater **20**(12), 1677–1682 (2021).

[71] L. Mogg, G.-P. Hao, S. Zhang, C. Bacaksiz, Y.-C. Zou, S.J. Haigh, F.M. Peeters, A.K. Geim, and M. Lozada-Hidalgo, "Atomically thin micas as proton-conducting membranes," Nat Nanotechnol **14**(10), 962–966 (2019).

[72] N.H. Jamil, and S. Palaniandy, "Acid medium sonication: A method for the preparation of low density talc nano-sheets," Powder Technol **200**(1–2), 87–90 (2010).

[73] S.M. Sousa, H.L.O. Morais, J.C.C. Santos, A.P.M. Barboza, B.R.A. Neves, E.S. Pinto, and M.C. Prado, "Liquid phase exfoliation of talc: effect of the medium on flake size and shape," Beilstein Journal of Nanotechnology **14**, 68–78 (2023).

[74] W. Wang, W. Zhen, S. Bian, and X. Xi, "Structure and properties of quaternary fulvic acid–intercalated saponite/poly(lactic acid) nanocomposites," Appl Clay Sci **109–110**, 136–142 (2015).

[75] R.A. Vaia, H. Ishii, and E.P. Giannelis, "Synthesis and properties of two-dimensional nanostructures by direct intercalation of polymer melts in layered silicates," Chemistry of Materials **5**(12), 1694–1696 (1993).

[76] S. Sinha Ray, and M. Okamoto, "Polymer/layered silicate nanocomposites: a review from preparation to processing," Prog Polym Sci **28**(11), 1539–1641 (2003).

[77] T.T. Zhu, C.H. Zhou, F.B. Kabwe, Q.Q. Wu, C.S. Li, and J.R. Zhang, "Exfoliation of montmorillonite and related properties of clay/polymer nanocomposites," Appl Clay Sci **169**, 48–66 (2019).

[78] R. Wang, H. Li, G. Ge, N. Dai, J. Rao, H. Ran, and Y. Zhang, "Montmorillonite-Based Two-Dimensional Nanocomposites: Preparation and Applications," Molecules **26**(9), 2521 (2021).

[79] J.T. Kloprogge, "Synthesis of Smectite Clay Minerals: A Critical Review," Clays Clay Miner **47**(5), 529–554 (1999).

[80] M. Claverie, A. Dumas, C. Carême, M. Poirier, C. Le Roux, P. Micoud, F. Martin, and C. Aymonier, "Synthetic Talc and Talc-Like Structures: Preparation, Features and Applications," Chemistry - A European Journal **24**(3), 519–542 (2018).

[81] T. Hammouda, M. Pichavant, P. Barbey, and A.J. Brearley, "Synthesis of fluorphlogopite single crystals. Applications to experimental studies," European Journal of Mineralogy **7**(6), 1381–1388 (1995).

[82] L.C. Becker, W.F. Bergfeld, D. V. Belsito, R.A. Hill, C.D. Klaassen, D.C. Liebler, J.G. Marks, R.C. Shank, T.J. Slaga, P.W. Snyder, F.A. Andersen, and L.J. Gill, "Safety Assessment of Synthetic Fluorphlogopite as Used in Cosmetics," Int J Toxicol **34**, 43S-52S (2015).

[83] L.B. de Paiva, A.R. Morales, and F.R. Valenzuela Díaz, "Organoclays: Properties, preparation and applications," Appl Clay Sci **42**(1–2), 8–24 (2008).

[84] B. Lebeau, J. Brendlé, C. Marichal, A.J. Patil, E. Muthusamy, and S. Mann, "One-Step Synthesis and Solvent-Induced Exfoliation of Hybrid Organic–Inorganic Phyllosilicate-Like Materials," J Nanosci Nanotechnol **6**(2), 352–359 (2006).

[85] S. Wang, H. Cao, Y. Zhong, Y. Yang, and Z. Shao, "A novel aminoclay–curcumin hybrid for enhanced chemotherapy," J Mater Chem B **4**(24), 4295–4301 (2016).

[86] S.Y. Kim, and Y.-S. Choi, "Preparation of magnesium-based two-dimensional phyllosilicate materials and simultaneous antioxidant drug intercalation," Colloids Surf A Physicochem Eng Asp **569**, 164–170 (2019).

[87] V.K.H. Bui, D. Park, and Y.-C. Lee, "Aminoclays for biological and environmental applications: An updated review," Chemical Engineering Journal **336**, 757–772 (2018).

[88] M. Tohidi Vahdat, K. Varoon Agrawal, and G. Pizzi, "Machine-learning accelerated identification of exfoliable two-dimensional materials," Mach Learn Sci Technol **3**(4), 045014 (2022).

[89] R. Mizuguchi, Y. Igarashi, H. Imai, and Y. Oaki, "Lateral-size control of exfoliated transition-metal–oxide 2D materials by machine learning on small data," Nanoscale **13**(6), 3853–3859 (2021).

[90] B. Tang, Y. Lu, J. Zhou, T. Chouhan, H. Wang, P. Golani, M. Xu, Q. Xu, C. Guan, and Z. Liu, "Machine learning-guided synthesis of advanced inorganic materials," Materials Today **41**, 72–80 (2020).

[91] E. Kim, K. Huang, A. Saunders, A. McCallum, G. Ceder, and E. Olivetti, "Materials Synthesis Insights from Scientific Literature via Text Extraction and Machine Learning," Chemistry of Materials **29**(21), 9436–9444 (2017).

[92] A. Wang, J.J. Freeman, and B.L. Jolliff, "Understanding the Raman spectral features of phyllosilicates," Journal of Raman Spectroscopy **46**(10), 829–845 (2015).

[93] C. Mosser, S. Petit, and M. Mestdagh, "ESR and IR Evidence for Chromium in Kaolinites," Clay Miner **28**(3), 353–364 (1993).

[94] V. Timón, C.S. Praveen, E. Escamilla-Roa, and M. Valant, "Hybrid density functional based study on the band structure of trioctahedral mica and its dependence on the variation of Fe2+ content," J Mol Graph Model **44**, 129–135 (2013).

[95] G.J. Borradaile, and T. Werner, "Magnetic anisotropy of some phyllosilicates," Tectonophysics **235**(3), 223–248 (1994).

[96] G. Ulian, and G. Valdrè, "Density functional investigation of the thermophysical and thermochemical properties of talc [Mg3Si4O10(OH)2]," Phys Chem Miner **42**(2), 151–162 (2014).

[97] G. Ulian, D. Moro, and G. Valdrè, "Infrared and Raman spectroscopic features of clinochlore Mg6Si4O10(OH)8: A density functional theory contribution," Appl Clay Sci **197**(July), 105779 (2020).

[98] G.R. Schleder, C.M. Acosta, and A. Fazzio, "Exploring Two-Dimensional Materials Thermodynamic Stability via Machine Learning," ACS Appl Mater Interfaces **12**(18), 20149–20157 (2020).

[99] S. Masubuchi, E. Watanabe, Y. Seo, S. Okazaki, T. Sasagawa, K. Watanabe, T. Taniguchi, and T. Machida, "Deep-learning-based image segmentation integrated with optical microscopy for automatically searching for two-dimensional materials," NPJ 2D Mater Appl **4**(1), 3 (2020).

[100] Y. Mao, L. Wang, C. Chen, Z. Yang, and J. Wang, "Thickness Determination of Ultrathin 2D Materials Empowered by Machine Learning Algorithms," Laser Photon Rev **17**(4), (2023).

[101] H. Yin, Z. Sun, Z. Wang, D. Tang, C.H. Pang, X. Yu, A.S. Barnard, H. Zhao, and Z. Yin, "The data-intensive scientific revolution occurring where two-dimensional materials meet machine learning," Cell Rep Phys Sci **2**(7), 100482 (2021).

[102] Y. Li, Y. Kong, J. Peng, C. Yu, Z. Li, P. Li, Y. Liu, C.-F. Gao, and R. Wu, "Rapid identification of two-dimensional materials via machine learning assisted optic microscopy," Journal of Materiomics **5**(3), 413–421 (2019).

[103] M. Ziatdinov, O. Dyck, A. Maksov, X. Li, X. Sang, K. Xiao, R.R. Unocic, R. Vasudevan, S. Jesse, and S. V. Kalinin, "Deep Learning of Atomically Resolved Scanning Transmission Electron Microscopy Images: Chemical Identification and Tracking Local Transformations," ACS Nano **11**(12), 12742–12752 (2017).

[104] N.C. Frey, D. Akinwande, D. Jariwala, and V.B. Shenoy, "Machine Learning-Enabled Design of Point Defects in 2D Materials for Quantum and Neuromorphic Information Processing," ACS Nano **14**(10), 13406–13417 (2020).

[105] M. Dou, and M. Fyta, "Lithium adsorption on 2D transition metal dichalcogenides: towards a descriptor for machine learned materials design," J Mater Chem A Mater **8**(44), 23511–23518 (2020).

[106] S.A. Tawfik, O. Isayev, C. Stampfl, J. Shapter, D.A. Winkler, and M.J. Ford, "Efficient Prediction of Structural and Electronic Properties of Hybrid 2D Materials Using Complementary DFT and Machine Learning Approaches," Adv Theory Simul **2**(1), 1800128 (2019).

[107] L. Bassman Oftelie, P. Rajak, R.K. Kalia, A. Nakano, F. Sha, J. Sun, D.J. Singh, M. Aykol, P. Huck, K. Persson, and P. Vashishta, "Active learning for accelerated design of layered materials," NPJ Comput Mater **4**(1), 74 (2018).

[108] G.A. Tritsaris, S. Carr, and G.R. Schleder, "Computational design of moiré assemblies aided by artificial intelligence," Appl Phys Rev **8**(3), 031401 (2021).

[109] B. Burger, P.M. Maffettone, V. V. Gusev, C.M. Aitchison, Y. Bai, X. Wang, X. Li, B.M. Alston, B. Li, R. Clowes, N. Rankin, B. Harris, R.S. Sprick, and A.I. Cooper, "A mobile robotic chemist," Nature **583**(7815), 237–241 (2020).

[110] L. Shafei, P. Adhikari, and W.-Y. Ching, "DFT Study of Electronic Structure and Optical Properties of Kaolinite, Muscovite, and Montmorillonite," Crystals (Basel) **11**(6), 618 (2021).

[111] D. Tunega, and A. Zaoui, "Mechanical and Bonding Behaviors Behind the Bending Mechanism of Kaolinite Clay Layers," The Journal of Physical Chemistry C **124**(13), 7432–7440 (2020).

[112] G. Ulian, D. Moro, and G. Valdrè, "First principle investigation of the mechanical properties of natural layered nanocomposite: Clinochlore as a model system for heterodesmic structures," Compos Struct **202**, 551–558 (2018).

[113] H. Chacham, A.P.M. Barboza, A.B. de Oliveira, C.K. de Oliveira, R.J.C. Batista, and B.R.A. Neves, "Universal deformation pathways and flexural hardening of nanoscale 2D-material standing folds," Nanotechnology **29**(9), 095704 (2018).

[114] R.J.C. Batista, R.F. Dias, A.P.M. Barboza, A.B. de Oliveira, T.M. Manhabosco, T.R. Gomes-Silva, M.J.S. Matos, A.C. Gadelha, C. Rabelo, L.G.L. Cançado, A. Jorio, H. Chacham, and B.R.A. Neves, "Nanomechanics of few-layer materials: do individual layers slide upon folding?," Beilstein Journal of Nanotechnology **11**, 1801–1808 (2020).

[115] C. Lee, X. Wei, J.W. Kysar, and J. Hone, "Measurement of the Elastic Properties and Intrinsic Strength of Monolayer Graphene," Science (1979) **321**(5887), 385–388 (2008).

[116] A. Castellanos-Gomez, M. Poot, G.A. Steele, H.S.J. Van Der Zant, N. Agraït, and G. Rubio-Bollinger, "Elastic properties of freely suspended MoS2 nanosheets," Advanced Materials **24**(6), 772–775 (2012).

[117] D.A. Kunz, E. Max, R. Weinkamer, T. Lunkenbein, J. Breu, and A. Fery, "Deformation Measurements on Thin Clay Tactoids," Small **5**(16), 1816–1820 (2009).

[118] G. Franceschi, P. Kocán, A. Conti, S. Brandstetter, J. Balajka, I. Sokolović, M. Valtiner, F. Mittendorfer, M. Schmid, M. Setvín, and U. Diebold, "Resolving the intrinsic short-range ordering of K+ ions on cleaved muscovite mica," Nat Commun **14**(1), 208 (2023).

[119] S. Miyake, and M. Wang, "Nanoprocessing of layered crystalline materials by atomic force microscopy," Nanoscale Res Lett **10**(1), 123 (2015).

[120] B. Ryu, L. Wang, H. Pu, M.K.Y. Chan, and J. Chen, "Understanding, discovery, and synthesis of 2D materials enabled by machine learning," Chem Soc Rev **51**(6), 1899–1925 (2022).

[121] X. Wang, D. Han, Y. Hong, H. Sun, J. Zhang, and J. Zhang, "Machine Learning Enabled Prediction of Mechanical Properties of Tungsten Disulfide Monolayer," ACS Omega **4**(6), 10121–10128 (2019).

[122] B. Sattari Baboukani, Z. Ye, K. G. Reyes, and P.C. Nalam, "Prediction of Nanoscale Friction for Two-Dimensional Materials Using a Machine Learning Approach," Tribol Lett **68**(2), 57 (2020).

[123] J.R. Petriglieri, E. Salvioli-Mariani, L. Mantovani, M. Tribaudino, P.P. Lottici, C. Laporte-Magoni, and D. Bersani, "Micro-Raman mapping of the polymorphs of serpentine," Journal of Raman Spectroscopy **46**(10), 953–958 (2015).

[124] A. Dumas, F. Martin, E. Ferrage, P. Micoud, C. Le Roux, and S. Petit, "Synthetic talc advances: Coming closer to nature, added value, and industrial requirements," Appl Clay Sci **85**(1), 8–18 (2013).

[125] J.J. Blaha, and G.J. Rosasco, "Raman Microprobe Spectra of Individual Microcrystals and Fibers of Talc, Tremolite, and Related Silicate Minerals," Anal Chem **50**(7), 892–896 (1978).

[126] C. Rinaudo, D. Gastaldi, and E. Belluso, "CHARACTERIZATION OF CHRYSOTILE, ANTIGORITE AND LIZARDITE BY FT-RAMAN SPECTROSCOPY," The Canadian Mineralogist **41**(4), 883–890 (2003).

[127] A. Tlili, D.C. Smith, J.-M. Beny, and H. Boyer, "A Raman microprobe study of natural micas," Mineral Mag **53**(370), 165–179 (1989).

[128] E. Loh, "Optical vibrations in sheet silicates," Journal of Physics C: Solid State Physics **6**(6), 1091–1104 (1973).

[129] M. Ishii, T. Shimanouchi, and M. Nakahira, "Far infra-red absorption spectra of layer silicates," Inorganica Chim Acta **1**, 387–392 (1967).

[130] V.C. Farmer, "The infra-red spectra of talc, saponite, and hectorite," Mineralogical Magazine and Journal of the Mineralogical Society **31**(241), 829–845 (1958).

[131] V.C. Farmer, and J.D. Russell, "The infra-red spectra of layer silicates," Spectrochimica Acta **20**(7), 1149–1173 (1964).

[132] E.W. Radoslovich, "The structure of muscovite, KAl2(Si3Al)O10(OH)2," Acta Crystallogr **13**(11), 919–932 (1960).

[133] V. STUBIČAN, and R. ROY, "A new approach to assignment of infra-red absorption bands in layer-structure silicates," Z Kristallogr Cryst Mater **115**(1–6), 200–214 (1961).

[134] S.N. Neal, H.-S. Kim, K.R. O'Neal, A. V. Haglund, K.A. Smith, D.G. Mandrus, H.A. Bechtel, G.L. Carr, K. Haule, D. Vanderbilt, and J.L. Musfeldt, "Symmetry crossover in layered MPS3 complexes (M=Mn, Fe, Ni) via near-field infrared spectroscopy," Phys Rev B **102**(8), 085408 (2020).

[135] S.N. Neal, K.R. O'Neal, A. Haglund, D.G. Mandrus, H. Bechtel, G.L. Carr, K. Haule, D. Vanderbilt, H.-S. Kim, and J.L. Musfeldt, "Exploring few and single layer CrPS 4 with near-field infrared spectroscopy," 2d Mater, 0–12 (2021).

[136] A.C. Gadelha, T.L. Vasconcelos, L.G. Cançado, and A. Jorio, "Nano-optical Imaging of In-Plane Homojunctions in Graphene and MoS 2 van der Waals Heterostructures on Talc and SiO 2," J Phys Chem Lett **12**(31), 7625–7631 (2021).

[137] I.D. Barcelos, H.A. Bechtel, C.J.S. de Matos, D.A. Bahamon, B. Kaestner, F.C.B. Maia, and R.O. Freitas, "Probing Polaritons in 2D Materials with Synchrotron Infrared Nanospectroscopy," Adv Opt Mater **8**(5), 1901091 (2020).

[138] R. Longuinhos, A. Vymazalová, A.R. Cabral, S.S. Alexandre, R.W. Nunes, and J. Ribeiro-Soares, "Raman spectrum of layered jacutingaite (Pt2HgSe3) crystals—Experimental and theoretical study," Journal of Raman Spectroscopy **51**(2), 357–365 (2020).

[139] R. Longuinhos, A. Vymazalová, A.R. Cabral, and J. Ribeiro-Soares, "Raman spectrum of layered tilkerodeite (Pd 2 HgSe 3 ) topological insulator: the palladium analogue of jacutingaite (Pt 2 HgSe 3 )," Journal of Physics: Condensed Matter **33**(6), 065401 (2021).

[140] F. Kargar, and A.A. Balandin, "Advances in Brillouin–Mandelstam light-scattering spectroscopy," Nat Photonics **15**(10), 720–731 (2021).

[141] S.A. Tawfik, M. Rashid, S. Gupta, S.P. Russo, T.R. Walsh, and S. Venkatesh, "Machine learning-based discovery of vibrationally stable materials," NPJ Comput Mater **9**(1), 5 (2023).

[142] B. Mortazavi, I.S. Novikov, E. V. Podryabinkin, S. Roche, T. Rabczuk, A. V. Shapeev, and X. Zhuang, "Exploring phononic properties of two-dimensional materials using machine learning interatomic potentials," Appl Mater Today **20**, 100685 (2020).

[143] D.A. Robinson, "Measurement of the Solid Dielectric Permittivity of Clay Minerals and Granular Samples Using a Time Domain Reflectometry Immersion Method," Vadose Zone Journal **3**(2), 705–713 (2004).

[144] A. Maruvada, K. Shubhakar, N. Raghavan, K.L. Pey, and S.J. O'Shea, "Dielectric breakdown of 2D muscovite mica," Sci Rep **12**(1), 14076 (2022).

[145] A. Arora, K.L. Ganapathi, T. Dixit, M. Miryala, M. Masato, M.S.R. Rao, and A. Krishnan, "Thickness-Dependent Nonlinear Electrical Conductivity of Few-Layer Muscovite Mica," Phys Rev Appl **17**(6), 064042 (2022).

[146] R. Zhang, W. Chen, C. Teng, W. Liao, B. Liu, and H.-M. Cheng, "Realization of a non-markov chain in a single 2D mineral RRAM," Sci Bull (Beijing) **66**(16), 1634–1640 (2021).

[147] Q. Ji, Y. Zhang, T. Gao, Y. Zhang, D. Ma, M. Liu, Y. Chen, X. Qiao, P.H. Tan, M. Kan, J. Feng, Q. Sun, and Z. Liu, "Epitaxial monolayer MoS2 on mica with novel photoluminescence," Nano Lett **13**(8), 3870–3877 (2013).

[148] F. Cui, C. Wang, X. Li, G. Wang, K. Liu, Z. Yang, Q. Feng, X. Liang, Z. Zhang, S. Liu, Z. Lei, Z. Liu, H. Xu, and J. Zhang, "Tellurium-Assisted Epitaxial Growth of Large-Area, Highly Crystalline ReS2 Atomic Layers on Mica Substrate," Advanced Materials **28**(25), 5019–5024 (2016).

[149] J. Su, M. Wang, Y. Li, F. Wang, Q. Chen, P. Luo, J. Han, S. Wang, H. Li, and T. Zhai, "Sub-Millimeter-Scale Monolayer p-Type H-Phase VS2," Adv Funct Mater **30**(17), 2000240 (2020).

[150] L. Deng, W. Li, J. Sun, X. Wang, Q. Zhang, C. Lin, K. Pan, Q. Yan, and S. Cheng, "Can fluorphlogopite mica be used as an alkali metal ion source to boost the growth of two-dimensional molybdenum dioxide?," Appl Surf Sci **612**, 155853 (2023).

[151] C. Wang, P. Guo, H. Jiang, J. Li, H. Zhu, J. Sun, X. Fan, L. Huang, and Y. Wang, "Application of Transparent Fluorphlogopite Substrate in Flexible Electromagnetic Devices," Adv Eng Mater **25**(6), 2201105 (2023).

[152] Y. He, H. Dong, Q. Meng, L. Jiang, W. Shao, L. He, and W. Hu, "Mica, a potential two-dimensional-crystal gate insulator for organic field-effect transistors," Advanced Materials **23**(46), 5502–5507 (2011).

[153] X. Zhang, Y. He, R. Li, H. Dong, and W. Hu, "2D Mica Crystal as Electret in Organic Field-Effect Transistors for Multistate Memory," Advanced Materials **28**(19), 3755–3760 (2016).

[154] G. Zhong, and J. Li, "Muscovite mica as a universal platform for flexible electronics," Journal of Materiomics **6**(2), 455–457 (2020).

[155] Y. Zhou, Y. Nie, Y. Liu, K. Yan, J. Hong, C. Jin, Y. Zhou, J. Yin, Z. Liu, and H. Peng, "Epitaxy and Photoresponse of Two-Dimensional GaSe Crystals on Flexible Transparent Mica Sheets," ACS Nano **8**(2), 1485–1490 (2014).

[156] J. Yin, Z. Tan, H. Hong, J. Wu, H. Yuan, Y. Liu, C. Chen, C. Tan, F. Yao, T. Li, Y. Chen, Z. Liu, K. Liu, and H. Peng, "Ultrafast and highly sensitive infrared photodetectors based on two-dimensional oxyselenide crystals," Nat Commun **9**(1), 3311 (2018).

[157] Y.-N. Lai, C.-H. Chang, P.-C. Wang, and Y.-H. Chu, "Highly efficient flexible organic light-emitting diodes based on a high-temperature durable mica substrate," Org Electron **75**, 105442 (2019).

[158] C.S. de Brito, C.R. Rabahi, M.D. Teodoro, D.F. Franco, M. Nalin, I.D. Barcelos, and Y.G. Gobato, "Strain engineering of quantum confinement in WSe2 on nano-roughness glass substrates," Appl Phys Lett **121**(7), 070601 (2022).

[159] F.S. Covre, P.E. Faria, V.O. Gordo, C.S. de Brito, Y. V. Zhumagulov, M.D. Teodoro, O.D.D. Couto, L. Misoguti, S. Pratavieira, M.B. Andrade, P.C.M. Christianen, J. Fabian, F. Withers, and Y. Galvão Gobato, "Revealing the impact of strain in the optical properties of bubbles in monolayer MoSe2," Nanoscale **14**(15), 5758–5768 (2022).

[160] F.J.R. Costa, T.G.-L. Brito, I.D. Barcelos, and L.F. Zagonel, "Impacts of dielectric screening on the luminescence of monolayer WSe2," Nanotechnology **34**(38), 385703 (2023).

[161] F. Shao, S.Y. Woo, N. Wu, R. Schneider, A.J. Mayne, S.M. de Vasconcellos, A. Arora, B.J. Carey, J.A. Preuß, N. Bonnet, M. Och, C. Mattevi, K. Watanabe, T. Taniguchi, Z. Niu, R. Bratschitsch, and L.H.G. Tizei, "Substrate influence on transition metal dichalcogenide monolayer exciton absorption linewidth broadening," Phys Rev Mater **6**(7), 074005 (2022).

[162] P.L. Mahapatra, A.K. Singh, R. Tromer, K. R., A. M., G. Costin, B. Lahiri, T.K. Kundu, P.M. Ajayan, E.I. Altman, D.S. Galvao, and C.S. Tiwary, "Energy harvesting using two-dimensional (2D) d-silicates from abundant natural minerals," J Mater Chem C Mater **11**(6), 2098–2106 (2023).

[163] P. Kumbhakar, C. Chowde Gowda, P.L. Mahapatra, M. Mukherjee, K.D. Malviya, M. Chaker, A. Chandra, B. Lahiri, P.M. Ajayan, D. Jariwala, A. Singh, and C.S. Tiwary, "Emerging 2D metal oxides and their applications," Materials Today **45**, 142–168 (2021).

[164] Y. Bai, H. Zhang, M. Zhang, D. Wang, H. Zeng, J. Zhao, H. Xue, G. Wu, J. Su, Y. Xie, Y. Zhang, H. Jing, H. Yu, Z. Hu, R. Peng, M. Wang, and Y. Wu, "Liquid-phase growth and optoelectronic properties of two-dimensional hybrid perovskites CH3NH3PbX3 (X = Cl, Br, I)," Nanoscale **12**(2), 1100–1108 (2020).

[165] Y. Xue, J. Yuan, J. Liu, and S. Li, "Controllable Synthesis of 2D Perovskite on Different Substrates and Its Application as Photodetector," Nanomaterials **8**(8), 591 (2018).

[166] C. Jia, X. Zhao, Y.-H. Lai, J. Zhao, P.-C. Wang, D.-S. Liou, P. Wang, Z. Liu, W. Zhang, W. Chen, Y.-H. Chu, and J. Li, "Highly flexible, robust, stable and high efficiency perovskite solar cells enabled by van der Waals epitaxy on mica substrate," Nano Energy **60**, 476–484 (2019).

[167] M. Gibertini, M. Koperski, A.F. Morpurgo, and K.S. Novoselov, "Magnetic 2D materials and heterostructures," Nat Nanotechnol **14**(5), 408–419 (2019).

[168] V.P. Ningrum, B. Liu, W. Wang, Y. Yin, Y. Cao, C. Zha, H. Xie, X. Jiang, Y. Sun, S. Qin, X. Chen, T. Qin, C. Zhu, L. Wang, and W. Huang, "Recent Advances in Two-Dimensional Magnets: Physics and Devices towards Spintronic Applications," Research **2020**, (2020).

[169] Q.H. Wang, A. Bedoya-Pinto, M. Blei, A.H. Dismukes, A. Hamo, S. Jenkins, M. Koperski, Y. Liu, Q.-C. Sun, E.J. Telford, H.H. Kim, M. Augustin, U. Vool, J.-X. Yin, L.H. Li, A. Falin, C.R. Dean, F. Casanova, R.F.L. Evans, M. Chshiev, A. Mishchenko, C. Petrovic, R. He, L. Zhao, A.W. Tsen, B.D. Gerardot, M. Brotons-Gisbert, Z. Guguchia, X. Roy, S. Tongay, Z. Wang, M.Z. Hasan, J. Wrachtrup, A. Yacoby, A. Fert, S. Parkin, K.S. Novoselov, P. Dai, L. Balicas, and E.J.G. Santos, "The Magnetic Genome of Two-Dimensional van der Waals Materials," ACS Nano **16**(5), 6960–7079 (2022).

[170] M. Onga, Y. Sugita, T. Ideue, Y. Nakagawa, R. Suzuki, Y. Motome, and Y. Iwasa, "Antiferromagnet–Semiconductor Van Der Waals Heterostructures: Interlayer Interplay of Exciton with Magnetic Ordering," Nano Lett **20**(6), 4625–4630 (2020).

[171] T. Norden, C. Zhao, P. Zhang, R. Sabirianov, A. Petrou, and H. Zeng, "Giant valley splitting in monolayer WS2 by magnetic proximity effect," Nat Commun **10**(1), 4163 (2019).

[172] T.P. Lyons, D. Gillard, A. Molina-Sánchez, A. Misra, F. Withers, P.S. Keatley, A. Kozikov, T. Taniguchi, K. Watanabe, K.S. Novoselov, J. Fernández-Rossier, and A.I. Tartakovskii, "Interplay between spin proximity effect and charge-dependent exciton dynamics in MoSe2/CrBr3 van der Waals heterostructures," Nat Commun **11**(1), 6021 (2020).

[173] D. Zhong, K.L. Seyler, X. Linpeng, N.P. Wilson, T. Taniguchi, K. Watanabe, M.A. McGuire, K.-M.C. Fu, D. Xiao, W. Yao, and X. Xu, "Layer-resolved magnetic proximity effect in van der Waals heterostructures," Nat Nanotechnol **15**(3), 187–191 (2020).

[174] K. Zollner, P.E. Faria Junior, and J. Fabian, "Proximity exchange effects in MoSe2 and WSe2 heterostructures with CrI3: Twist angle, layer, and gate dependence," Phys Rev B **100**(8), 085128 (2019).
[175] A.N. Grigorenko, M. Polini, and K.S. Novoselov, "Graphene plasmonics," Nat Photonics **6**(11), 749–758 (2012).
[176] F.C.B. Maia, B.T. O'Callahan, A.R. Cadore, I.D. Barcelos, L.C. Campos, K. Watanabe, T. Taniguchi, C. Deneke, A. Belyanin, M.B. Raschke, and R.O. Freitas, "Anisotropic Flow Control and Gate Modulation of Hybrid Phonon-Polaritons," Nano Lett **19**(2), 708–715 (2019).
[177] F.H. Feres, I.D. Barcelos, A.R. Cadore, L. Wehmeier, T. Nörenberg, R.A. Mayer, R.O. Freitas, L.M. Eng, S.C. Kehr, and F.C.B. Maia, "Graphene Nano-Optics in the Terahertz Gap," Nano Lett **23**(9), 3913–3920 (2023).
[178] F.H.L. Koppens, D.E. Chang, and F.J. García de Abajo, "Graphene Plasmonics: A Platform for Strong Light–Matter Interactions," Nano Lett **11**(8), 3370–3377 (2011).
[179] H. Hu, X. Guo, D. Hu, Z. Sun, X. Yang, and Q. Dai, "Flexible and Electrically Tunable Plasmons in Graphene-Mica Heterostructures," Advanced Science **5**(8), 1800175 (2018).
[180] A. Fali, S. Gamage, M. Howard, T.G. Folland, N.A. Mahadik, T. Tiwald, K. Bolotin, J.D. Caldwell, and Y. Abate, "Nanoscale Spectroscopy of Dielectric Properties of Mica," ACS Photonics **8**(1), 175–181 (2021).
[181] Y. Zhang, W. Xu, G. Liu, Z. Zhang, J. Zhu, and M. Li, "Bandgap prediction of two-dimensional materials using machine learning," PLoS One **16**(8), e0255637 (2021).
[182] A.C. Rajan, A. Mishra, S. Satsangi, R. Vaish, H. Mizuseki, K.-R. Lee, and A.K. Singh, "Machine-Learning-Assisted Accurate Band Gap Predictions of Functionalized MXene," Chemistry of Materials **30**(12), 4031–4038 (2018).
[183] H. Jin, H. Zhang, J. Li, T. Wang, L. Wan, H. Guo, and Y. Wei, "Discovery of Novel Two-Dimensional Photovoltaic Materials Accelerated by Machine Learning," J Phys Chem Lett **11**(8), 3075–3081 (2020).
[184] L. Ge, H. Yuan, Y. Min, L. Li, S. Chen, L. Xu, and W.A. Goddard, "Predicted Optimal Bifunctional Electrocatalysts for the Hydrogen Evolution Reaction and the Oxygen Evolution Reaction Using Chalcogenide Heterostructures Based on Machine Learning Analysis of in Silico Quantum Mechanics Based High Throughput Screening," J Phys Chem Lett **11**(3), 869–876 (2020).
[185] K. Choudhary, M. Bercx, J. Jiang, R. Pachter, D. Lamoen, and F. Tavazza, "Accelerated Discovery of Efficient Solar Cell Materials Using Quantum and Machine-Learning Methods," Chemistry of Materials **31**(15), 5900–5908 (2019).
[186] Q. Chen, M. Chen, L. Zhu, N. Miao, J. Zhou, G.J. Ackland, and Z. Sun, "Composition-Gradient-Mediated Semiconductor–Metal Transition in Ternary Transition-Metal-Dichalcogenide Bilayers," ACS Appl Mater Interfaces **12**(40), 45184–45191 (2020).
[187] S.M. Neumayer, M.A. Susner, M.A. McGuire, S.T. Pantelides, S. Kalnaus, P. Maksymovych, and N. Balke, "Lowering of Tc in Van Der Waals Layered Materials Under In-Plane Strain," IEEE Trans Ultrason Ferroelectr Freq Control **68**(2), 253–258 (2021).
[188] C.M. Acosta, E. Ogoshi, J.A. Souza, and G.M. Dalpian, "Machine Learning Study of the Magnetic Ordering in 2D Materials," ACS Appl Mater Interfaces **14**(7), 9418–9432 (2022).
[189] G.R. Schleder, B. Focassio, and A. Fazzio, "Machine learning for materials discovery: Two-dimensional topological insulators," Appl Phys Rev **8**(3), 031409 (2021).
[190] K. Tanaka, K. Hachiya, W. Zhang, K. Matsuda, and Y. Miyauchi, "Machine-Learning Analysis to Predict the Exciton Valley Polarization Landscape of 2D Semiconductors," ACS Nano **13**(11), 12687–12693 (2019).
[191] M.F. Brigatti, M. Affronte, C. Elmi, D. Malferrari, and A. Laurora, "Trioctahedral Fe-rich micas: Relationships between magnetic behavior and crystal chemistry," American Mineralogist **100**(10), 2231–2241 (2015).
[192] S. Pini, M.F. Brigatti, M. Affronte, D. Malferrari, and A. Marcelli, "Magnetic behavior of trioctahedral micas with different octahedral Fe ordering," Phys Chem Miner **39**(8), 665–674 (2012).
[193] J.M.D. Coey, A. Moukarika, and O. Ballet, "Magnetic order in silicate minerals (invited)," J Appl Phys **53**(11), 8320–8325 (1982).
[194] K. Saritas, N. Doudin, E.I. Altman, and S. Ismail-Beigi, "Magnetism and piezoelectricity in stable transition metal silicate monolayers," Phys Rev Mater **5**(10), 104002 (2021).
[195] A. Chandrasekaran, D. Kamal, R. Batra, C. Kim, L. Chen, and R. Ramprasad, "Solving the electronic structure problem with machine learning," NPJ Comput Mater **5**(1), 22 (2019).
[196] S. Bhowmik, and A. Govind Rajan, "Chemical vapor deposition of 2D materials: A review of modeling, simulation, and machine learning studies," IScience **25**(3), 103832 (2022).
[197] M.E. Marchwiany, M. Birowska, M. Popielski, J.A. Majewski, and A.M. Jastrzębska, "Surface-Related Features Responsible for Cytotoxic Behavior of MXenes Layered Materials Predicted with Machine Learning Approach," Materials **13**(14), 3083 (2020).
[198] K. Barry, G.L. Lecomte-Nana, M. Seynou, M. Faucher, P. Blanchart, and C. Peyratout, "Comparative Properties of Porous Phyllosilicate-Based Ceramics Shaped by Freeze-Tape Casting," Ceramics **5**(1), 75–96 (2022).